\documentclass[useAMS,fleqn,usenatbib]{mnras}
\usepackage[T1]{fontenc}
\usepackage{ae,aecompl}
\usepackage{bethmacros}
\usepackage{graphicx}
\usepackage{amssymb}
\usepackage{amsmath}
\hypersetup{draft}
\def\spose#1{\hbox to 0pt{#1\hss}}
\def\lta{\mathrel{\spose{\lower 3pt\hbox{$\mathchar"218$}}
     \raise 2.0pt\hbox{$\mathchar"13C$}}}
\def\gta{\mathrel{\spose{\lower 3pt\hbox{$\mathchar"218$}}
     \raise 2.0pt\hbox{$\mathchar"13E$}}}

\title[Galactic Chaos]{Chaos and Variance in Galaxy Formation}

\author[Keller et al.]{B.\,W. Keller$^1$\thanks{Email: benjamin.keller `at'
uni-heidelberg.de},  J. W. Wadsley$^2$,  L. Wang$^3$, J. M. Diederik Kruijssen$^1$
\vspace*{6pt}\\
$^1$Zentrum f{\"u}r Astronomie der Universit\"at Heidelberg, Astronomisches
Rechen-Institut,  M{\"o}nchofstra{\ss}e 12-14, D-69120 Heidelberg, Germany\\
$^2$Department of Physics and Astronomy, McMaster University, Hamilton, Ontario,
L8S 4M1, Canada\\
$^3$International Centre For Radio Astronomy Research (ICRAR), University of
Western Australia, Crawley, WA 6009, Australia
}

\begin{document}
\maketitle
\label{firstpage}
\begin{abstract}
The evolution of galaxies is governed by equations with chaotic solutions:
gravity and compressible (magneto-)hydrodynamics. While this micro-scale
chaos and stochasticity has been well studied, it is poorly understood how
it couples to macro-scale properties examined in simulations of galaxy
formation. In this paper, we use tiny perturbations introduced by
floating-point roundoff, random number generators, and seemingly trivial
differences in algorithmic behaviour as seeds for chaotic behaviour.
These can ultimately grow to produce non-trivial differences in
star formation histories, circumgalactic medium (CGM) properties, and the
distribution of stellar mass.  We examine the importance of stochasticity
due to discreteness noise, variations in merger timings and how
self-regulation moderates the effects of this stochasticity. We show that
chaotic variations in stellar mass can be suppressed through gas exhaustion,
or be maintained at a roughly constant variance through stellar feedback. We
also find that galaxy mergers are critical points from which large (as much
as a factor of 2) variations can grow in global quantities such as the total
stellar mass.  These variations can grow and persist for Gyrs before
regressing towards the mean. These results show that detailed comparisons of
simulations require serious consideration of the magnitude of effects
compared to run-to-run chaotic variation, and may significantly complicate
interpreting the impact of different physical models.  Understanding the
results of simulations requires us to understand that the process of
simulation is not a mapping of an infinitesimal point in configuration space
to another, final infinitesimal point. Instead, simulations map a point in a
space of possible initial conditions points to a volume of possible final
states. 
\end{abstract}


\section{Introduction}
Numerical simulations of galaxy formation seek to map a set of initial
conditions to a final state, using (often phenomenological) models of the
physics involved. By comparing the evolution of identical initial conditions
with models based on differing physical assumptions, we can learn how these
assumptions impact the large-scale behaviour of the galaxy.  We then hope to
determine which of those assumptions are correct by comparing to
empirical evidence provided by observational studies. This relies on a primary
assumption: that all the differences between two simulated galaxies generated
with different models can be explained by the different physical processes
captured by these models. This assumption is incorrect.

It has been known for decades that collisionless N-body systems exhibit
irreversible, chaotic behaviour \citep{Miller1964,Kandrup1991}. While this is
important for tracking the individual orbits of planets in a solar system
\citep{Laskar1994} or stars within a cluster \citep{Miller1964}, it has
generally been regarded as unimportant to the global evolution of galaxy-scale
properties. Although collisionless dynamics in galactic potentials feature
chaotic orbits \citep{Henon1964,Pfenninger2009}, this feature has not been
considered particularly important to the overall assembly and star formation
history of the galaxy as a whole. In this paper, we show that the chaotic nature
of N-body systems, especially when coupled to collisional gas and subgrid models
for star formation and stellar feedback, limits the ability to interpret
simulation studies of galaxy formation and evolution. This chaotic behaviour
cannot be overcome with more sophisticated algorithms or higher resolution: it
is a fundamental feature of the physics governing galaxy formation.

Perhaps even more critical than N-body chaos is the stochastic, chaotic
behaviour of both compressible hydrodynamic and magnetohydrodynamic (MHD)
turbulence.  It has long been known that the structure of the interstellar
medium (ISM) is highly inhomogeneous \citep{Larson1981}.  With a Reynolds
numbers of $10^5-10^7$, the ISM should be spontaneously turbulent
\citep{Elmegreen2004}, and thus produce naturally the inhomogeneities we see.
As \citet{Ottino1989} identifies, compressible turbulence shares the
quasi-ergodic properties of classical dynamical chaos, but in a much higher
dimensional space.  Turbulent ISM overdensities can collapse to form stars
\citep{Larson1981}, which can then inject further energy into the turbulent
fluid through stellar feedback.  These features mean that different physical
scales within the ISM are strongly coupled to one another, and that small-scale
stochasticity can quickly ``pump'' large-scale variations.  Adding to this is
the fact that the ISM is magnetized \citep{Beck1996} and that magnetized
turbulence can be driven spontaneously through MHD fluid instabilities
\citep{Cho2003,Eyink2011}, the ISM has many routes through which its dynamics
may become chaotic.  The question of whether these internal chaotic dynamics
impact large-scale, galaxy averaged properties has yet to be addressed.

Many studies have looked at the difference between hydrodynamical schemes
\citep{Frenk1999,Agertz2007,Vazza2011,Kim2016}, models for star formation
\citep{Hopkins2011,Benincasa2016} or feedback
\citep{Thacker2000,Wurster2013,Keller2015}, or combinations of these
\citep{Nagai2007,Scannapieco2012}. In general, the consensus seems to be
emerging that sub-resolution physics models (i.e. those models used for
radiative cooling, star formation, feedback, etc.) are the primary source of
variation between models of galaxy formation
\citep{Scannapieco2012,Stinson2013,Su2017}.  Trajectories of individual
resolution elements are almost never considered in these studies, for exactly
the reasons above: they are well-known to be chaotic, and understanding the
causal chain that results in diverging trajectories is infeasible.  Instead, we
look at globally averaged properties: stellar masses, star formation rates,
density PDFs, etc.  However, these properties are ultimately a consequence of
the trajectories followed by the components of a simulated galaxy. 

The standard toolkit for studying chaotic behaviour in N-body systems does not
necessarily lend itself easily to studying large-scale properties of galaxies.
Typically, the onset of chaos can be quantified by looking at the Lyapunov
exponent, $\lambda$. Two particles, separated by an infinitesimal initial
distance
$\delta\mathbf{X_0}$ in phase space find their phase space distance $\mathbf{X}$
grow exponentially:
\begin{equation}
    \mathbf{X}(t) = e^{\lambda t}\delta\mathbf{X_0}
\end{equation}
The inverse of this ($\lambda^{-1}$) defines the characteristic e-folding time
for the divergence of chaotic trajectories.  For global galaxy properties,
unlike the trajectories of individual particles, there is no clear way to define
a phase-space metric, but we can simply look at 1-dimensional changes
in single global galaxy properties. Microscopic chaotic behaviour is
typically assumed to average out in some statistical sense. However, even
massive galaxies are not large enough systems to consider them in the
thermodynamic limit. Our own Galaxy contains  a few thousand star-forming
molecular clouds \citep{Rice2016,Miville2017}, a comparable number of young
stellar clusters \citep{Kharchenko2013}, and just over 150 globular clusters
\citep[2010 edition]{Harris2010}.  The hierarchical assembly of galaxies over cosmic time is
the critical place where these few-N effects can manifest.  Mergers occur
between only a handful of objects, and each merger can subsequently affect the
progression of the next.  Gravitational, radiative, and hydrodynamic
instabilities can all act to pump small-scale perturbations up to macroscopic
changes. On the other hand, conservation laws constrain how large these
variations can grow (after all, a galaxy can only form as many stars as it has
baryons). Beyond simple conservation laws, galaxy evolution is known to be
governed by a number of self-regulating processes
\citep{Silk1997,Ostriker2010,Benincasa2016}: these
too should act to damp down any perturbations that occur. Effectively,
galaxy-scale chaotic variations are constrained by physical self-regulation and
regression to the mean. 

The effect of chaos on collisionless systems has been well studied. It has been
demonstrated by \citet{Sellwood2009} that different random seeds used to
generate ``identical'' disc profiles can produce galaxies with a remarkably wide
range of bar amplitudes and pattern speeds. In fact, this even applies for truly
identical particle positions in their initial conditions, as they showed by
reflecting a simple N-body disc galaxy about an axis, introducing different
roundoff errors from its twin, unreflected disc. They found that the
\citet{Miller1964} instability produces macroscopic differences in bar amplitude
and pattern speed with a Lyapunov time 4 times shorter than the orbital period
at $2.5R_d$. For a galaxy like our own, that means nearly 280 e-foldings can
take place in a Hubble time, more than enough time for perturbations to become
non-linear.  \citet{Benhaiem2018} have also recently observed
similar macroscopic variations in the shape parameter of collisionless,
virialized halos. 

Bar structures are important for bulge formation
\citep{Pfenninger1990,Masters2011}, AGN fueling \citep{Noguchi1988,Laine2002},
and may even help form dark matter cores \citep{Weinberg2002}. This
stochasticity may therefore be coupled to many other galaxy properties. On
larger scales, \citet{Thiebaut2008} demonstrated that the chaotic behaviour is
limited to non-linear structures on scales below $3.5\Mpc\; h^{-1}$. Within halos,
however, the positions and even the number of substructures can vary significantly
due to small perturbations in the initial Gaussian random field from which they
formed. Halo mass and spins are generally robust to small perturbations, but
the direction of spin is not. A change in the substructure (and thus merger
history), as well as the orientation of the disc relative to these structures,
would likely have a non-negligible impact on the overall evolution of the galaxy
within this halo.  When coupled with hydrodynamics, star formation, and stellar
feedback, these small perturbations may grow to significant differences in the
evolution of ISM, the star formation history, and the distribution of metals.

The astrophysical discipline that has put the most thought into chaos and
reproducibility (when it comes to the interpretation of simulation results and
actually calculating Lyapunov coefficients) is likely that of the formation and
evolution of planetary systems. An excellent, if somewhat unsettling, example of
this are the studies that have found our own solar system is dynamically
unstable \citep{Laskar1994,Batygin2008,Laskar2009}.  As \citet{Batygin2008}
showed, the Lyapunov exponents for all of the terrestrial planets are all on the
order of $10^{-7}\;\mathrm{yr^{-1}}$, meaning that the planets will experience
hundreds of Lyapunov times in the remaining lifespan of the solar system.
Disturbingly, \citet{Laskar2009} showed that perturbing the position of Mercury
by $0.38\;\mathrm{mm}$(!) can put it on a collisional path with the Earth in one
of 2501 different realizations of these perturbed systems.  This chaos also
manifests itself in simulations of planet formation, as \citet{Hoffmann2017}
showed. They found that identical initial conditions, run with an identical
code, could still produce vastly different solar systems simply due to different
floating-point roundoff errors introduced by nondeterministic parallel
reductions, amplified by chaotic N-body interactions, resulting in a different
collisional history.

Recently, a pair of papers \citep{Su2018,Su2017} have used the stochasticity of
galaxy evolution to constrain the importance of different physical mechanisms.
\citet{Su2017} modified their feedback coupling scheme, ``re-shuffling''
feedback energy and momentum, randomizing which neighbours of a star particle
receive which fraction of the ejecta for a set of 5 runs. \citet{Su2018} simply
changed the random number seed used for 2 runs. \citet{Su2017} found little
variation in the final stellar mass for their $1.2\times10^{12}\Msun$ halo, with
significant noise in the instantaneous star formation rate. \citet{Su2018}
finds, for a $8\times10^9\Msun$ halo, a factor of $\sim 2$ variation in stellar
mass formed, as well as order-of-magnitude instantaneous variation in outflow
and star formation rates (though both of these values are quite noisy). 

In this study, we examine the numerical sources of microscopic perturbations
that can drive ``identical'' simulations to produce macroscopically divergent
results.  We show that these perturbations can arise in any code (by, for
example, changing the seed of a random number generator or simply through
asynchronous MPI reductions) and can result in divergent star formation
histories under certain conditions.  Our results show that macroscopic
differences in the stellar masses, stellar disc profiles, and CGM properties are
common in identical simulations of cosmological galaxies, and can persist for a
significant fraction of a Hubble time.  These results also show that, under the
influence of stellar feedback, chaotic variations can be pushed towards a
variances much higher than expected by Poisson noise alone, independent of galaxy
properties or even the details of feedback implementation.

\section{Numerical Methods Can Seed Perturbations}
In general, simulations can introduce perturbations into ``identical'' initial
conditions through three sources: discretization of the initial conditions
\citep{Romeo2003,Romeo2008,vandenBosch2018}, the use of random number generators
(RNGs) in sub-resolution models, and floating- point roundoff. This study will
focus on perturbations generated by the last two, but the results are general to
any source of small perturbations.  These perturbations are especially important
because they will arise in comparison studies that use identical initial
conditions with different simulation models.

Variations are also clearly introduced by algorithmic differences.  A change to
the method of solving the hydrodynamic or N-body equations will of course lead
to different (hopefully small) error terms, which like the perturbations
introduced by the previous sources, may grow exponentially due to the formally
chaotic behaviour of N-body systems.  Algorithmic differences can also occur
with identical codes run on different hardware, such as a running with a larger
or smaller number of cores.  Changing the number of cores that a code is run on
will change how the domain is decomposed, and this can lead to differences both
in the error terms and in the floating-point roundoff behaviour.  This means
that the variations explored in this paper act as an unavoidable floor to the
ability to both compare simulations run with different physics, simulation
codes, or even to reproduce past simulations on different hardware. Reproducible
codes do exist \citep{Springel2010}, which produce bitwise identical results,
but in these cases, these codes merely fix as constant the error terms
introduced by floating-point roundoff.  These errors still exist, however, and
still perturb the solutions.  To explore the range of variation in possible
outcomes, such codes require explicit perturbations in
initial conditions (as is done in \citealt{Genel2018}) or in parameters (as is
done in \citealt{Su2017,Su2018}).  If any additional perturbation is applied,
such as changing parameters or sub-grid models, all codes will produce different
roundoff-level perturbations, uncorrelated with previous runs, and unrelated to
the physical differences between the parameters or sub-grid model.
Reproducible codes are not more reliable predictors of physical outcomes, but
are simply easier to debug.

\subsection{Random Number Generators}\label{RNG}
Naturally, any sub-grid model that relies on RNGs
(e.g. Monte-Carlo integrators) will have a source of stochasticity in that RNG.
It is extremely common in astrophysical codes to use RNGs to convert rate
equations into a probability-per-timestep, and then draw a random number to
decide whether to apply that change or not
\citep[e.g.][]{Katz1992,Springel2005,DallaVecchia2008,Keller2014}. The ubiquitous
\citet{Katz1992} star formation algorithm transforms the star formation rate
equation $\dot \rho_* = c_* \rho_{\rm gas} / t_{\rm ff}$ into a probability of forming a
star particle per timestep $\Delta t$:
\begin{equation}\label{katz_sf_prob} 
    \mathrm{Pr} = 1 - e^{-c_*\Delta t/t_{\rm ff}}
\end{equation} 
Whether a star particle is formed on a given timestep is then determined
by drawing a random number which is compared to this probability. Thus, two
realizations of the same simulation can produce different behaviour due to a
different set of random numbers being drawn by the RNG. These can act as seed
perturbations, as a gas particle that forms a star particle becomes decoupled from
hydrodynamic forces, and can impart energy, momentum, and metals to the surrounding
medium if feedback is included in a simulation.

Star formation is not the only process that RNGs are used to model. Many
feedback models also rely on RNGs. The popular \citet{Springel2003} multiphase
model for galactic wind formation relies on a RNG both to select whether a
particle is given a momentum kick, but also for selecting the direction of that
momentum kick (as do the kinetic feedback models of \citealt{DallaVecchia2008}
and \citealt{Hopkins2011}). \citet{DallaVecchia2012} derive a stochastic
formulation for energy injection that ensures a constant post-feedback gas
temperature, which gives a probability for a gas particle to receive feedback
energy inversely proportional to this chosen temperature. 

\subsection{Floating-Point Arithmetic}\label{floating_point}
Removing all RNG-based subgrid models from a simulation will not necessarily banish
run-to-run variation. Arithmetic on floating-point numbers is non-associative:
{\tt a + (b + c)} is not necessarily equal to {\tt (a + b) + c}
\citep{Coonen1979}. While this may seem trivial, considering that
double-precision roundoff errors are typically on the order of one part in
$10^{15}$, as long as the Lyapunov time for the system being evolved is
sufficiently small, even these minuscule errors can be amplified to become
macroscopic changes in the final state of the system. Not only does this mean
that simply changing the order of operations in the source code of a simulation
software can drastically change the output of that simulation code, it can also
cause bitwise-identical compiled executables to produce different results in
certain cases.

It has been well documented in the computer science literature that on parallel
systems, reduction operations (such as summations) are non-deterministic
\citep{He2001,Diethelm2012}. Variations in network congestion, cache space, and
other subtle, unfortunately non-idealized physical processes involved in machines
running parallel reductions will introduce floating-point errors due to a change
in the order of associative operations. While work has been done to propose
deterministic, reproducible algorithms for reduction operations
\citep{Balaji2013}, these algorithms are still in the experimental stages, and
have not become the default for commonly used parallel protocols such as {\sc
MPI} or {\sc OpenMP}. These variations in roundoff error are the source of
perturbations used in \citet{Hoffmann2017}, and as they show, they can
ultimately cause significant variations in astrophysical simulations.

Even in the case of single threaded applications, or applications where parallel
reductions are avoided, different compiler optimization options can potentially
introduce differences on the order of the floating-point roundoff. While the {\sc
C99} (and later) standard guarantees that all optimization flags up to {\tt -O3} will
produce bitwise-identical floating-point operations, using the {\tt -Ofast} or
{\tt -ffast-math} compiler optimizations with {\tt gcc} will not produce
bitwise-identical results when comparing to compilations made without those flags. The
Intel C Compiler, {\tt icc} enables equivalent features by default.

Even if all of these issues are avoided, most large-scale simulations of galaxy
evolution rely on checkpointing: periodically writing out the state of the
simulation so that in the event of a hardware failure or a runtime longer than
allowed by a queuing system the simulation can be restarted.  This process of
writing out the contents of memory and then re-reading them again can introduce
subtle changes in the order of operations of a simulation, and again introduce
floating-point roundoff errors that can lead to different ``identical''
simulations diverging based on when or if they resumed from a checkpoint.

\section{Numerical Methods}
We study how small, numerical perturbations grow by running multiple
realizations of the
same initial condition on the same machine, with the same number of cores, using
the same compiled binary of our simulation codes.
\subsection{Initial Conditions}
\subsubsection{MUGS2 Cosmological Galaxy}
The cosmological zoom-in simulation used in this study is a member of the
McMaster Unbiased Galaxy Sample 2 (MUGS2) \citep{Keller2016}. The MUGS2 initial
conditions (ICs) were initially described in \citet{Stinson2010}. The zoom-in
ICs were selected from a $50h^{-1}\Mpc$ pure N-body cube, using a $WMAP3$
$\Lambda$CDM cosmology, with $H_0=73\kms\Mpc^{-1}$, $\Omega_M=0.24$,
$\Omega_{\rm bary}=0.04$, $\Omega_\Lambda=0.76$, and $\sigma_8=0.76$. The MUGS2
simulations have a gas particle mass of $2.2\times10^{5}\Msun$, and a
gravitational softening length of $312.5\pc$. A floor is applied to the SPH
smoothing length, ensuring it is never below 1/4 this value.  The star
formation threshold used for this galaxy is $9.3\;\rm{cm^{-3}}$, as was used in
previous studies \citep{Stinson2013,Keller2015,Keller2016}. The particular
galaxy we have used here is g1536, a halo with a virial mass of
$6.5\times10^{11}\Msun$. In the fiducial MUGS2 run, g1536 forms
$1.9\times10^{10}\Msun$ of stars by $z=0$. This particular halo has been used in a number
of studies beyond MUGS and MUGS2 \citep{Calura2012,Obreja2013,Nickerson2013},
including the MaGICC simulations \citep{Stinson2013}. 

\subsubsection{Isolated Dwarf}
We also ran a series of simulations using a modified version of the AGORA
\citep{Kim2014} isolated Milky Way-like Galaxy. The AGORA comparison project has
produced a common set of initial conditions for both cosmological zoom-in
simulations as well as isolated galaxies, and used these initial conditions to
compare 9 different simulation codes \citep{Kim2016}. The details of the IC and
how it was created can be found in \citet{Kim2014}. We modify the AGORA Milky
Way to match the ``isolated dwarf'' case used in \citet{Keller2014} in order to
increase the gas depletion time when simulated without feedback. As in
\citet{Keller2014}, we scaled down masses by a factor of 100, and lengths by
$100^{1/3}$ as compared to the Milky Way initial conditions, preserving physical
densities while lowering the gas surface densities by a factor of $\sim4.5$.
The gas particle mass for this dwarf IC is thus $850\Msun$, and the softening
lengths are set to $4.3\pc$. The IC has a total gas mass of
$8.6\times10^7\Msun$, and a stellar mass of $3.43\times10^8\Msun$. The disc has
a scale length of $740\pc$, a gas fraction of 0.2, and a scale height of
$74\pc$. It is embedded within a halo with $M_{200} = 1.074\times10^{10}\Msun$,
with corresponding concentration parameter $c=10$ and spin parameter $\lambda =
0.04$.  With this initial condition, we used a star formation density threshold
of $50\; \rm{cm^{-3}}$.

\subsection{Simulation Codes}
To show that the effects we study here are independent of subgrid physics,
gravitational solver, or hydrodynamics methods, we use two different simulation
codes along with a variety of feedback algorithms.  The two codes we use, {\sc
Gasoline2} and {\sc RAMSES}, are Lagrangian and Eulerian (respectively)
astrophysical simulation codes frequently used in simulations of galaxy
formation. Both {\sc Gasoline2} and {\sc RAMSES} were modified to ensure that
all random number generators were seeded with the same value for each run (0 in
this case).

\subsubsection{{\sc Gasoline2}}
{\sc Gasoline2} \citep{Wadsley2017} is a modern update to the {\sc Gasoline}
\citep{Wadsley2004} parallel TreeSPH \citep{Katz1996} simulation code. {\sc
Gasoline} pairs a \citet{Barnes1986} KD-Tree method for solving Poisson's
equation, along with the Lagrangian Smoothed Particle Hydrodynamics (SPH)
\citep{Gingold1977} method for solving Euler's equations. {\sc Gasoline2}
includes a number of updates that improve the performance and accuracy of the
hydrodynamics method originally used in {\sc Gasoline}. These features include a
\citet{Saitoh2009}-style timestep limiter, Wendland C2 and C4 smoothing kernels,
an artificial diffusion term first described in \citet{Wadsley2008}, and a
unique geometrically averaged density force \citep{Wadsley2017}. These
improvements allow {\sc Gasoline2} to accurately handle problems involving
strong shocks, multiphase flows, and problems involving mixing. This effectively
eliminates the problems seen in so-called ``traditional'' SPH methods
\citep{Ritchie2001,Agertz2007,Read2010}.  {\sc Gasoline} has been used for
hundreds of studies of galaxy formation, and {\sc Gasoline2} has been used by
dozens of studies already.

\subsubsection{{\sc RAMSES}}
{\sc RAMSES} \citep{Teyssier2002} is an Eulerian, Adaptive Mesh Refinement
(AMR) N-body and hydrodynamical code. AMR allows for spatial adaptivity, with
higher resolution in regions (in this study) of increased density (the
so-called ``quasi-Lagrangian'' mode). {\sc RAMSES} solves the compressible
Euler equations by means of an unsplit second-order Monotone Upstream-centered
Scheme for Conservation Laws (MUSCL) Godunov method \citep{vanLeer1979}, with
an HLLC Riemann Solver. Poisson's equation is solved for gravity using a
``one-way interface'' scheme \citep{Jessop1994}. {\sc RAMSES} includes physics
for radiative cooling, star formation, and stellar feedback, which we describe
in section~\ref{sg-phys}. Like {\sc Gasoline}, {\sc RAMSES} has been used for
hundreds of studies of cosmological structure and galaxy formation.

\subsection{Subgrid Physics}\label{sg-phys}
\subsubsection{Star Formation}
{\sc Gasoline2} and {\sc RAMSES} both adopt a \citet{Katz1992} style star
formation model. The star formation rate is calculated to be proportional to a
ratio of the gas density to the free-fall time, with some rate coefficient $c_*
< 1$. In {\sc Gasoline2} this is converted to a probability using
equation~\ref{katz_sf_prob}. {\sc RAMSES}, being an Eulerian code, has 
grid cells with varying masses. In order to avoid spurious (de-)refinement by
star formation, it specifies a constant stellar mass $m_*$. A star-forming cell
thus has a probability of forming $N$ star particles of equal mass, given by the
following equation \citep{Rasera2006}:
\begin{equation}\label{ramses_sf_prob}
	\mathrm{Pr}(N) = \frac{\lambda^N}{N!}\exp{(-\lambda)}
\end{equation}
Where the parameter $\lambda$ is set by the cell resolution and stellar mass:
\begin{equation}\label{ramses_sf_lambda}
    \lambda = \frac{\rho \Delta x^3}{m_*}\frac{c* \Delta t}{t_{\rm ff}}
\end{equation}

Both codes also rely on a maximum temperature and minimum density threshold to
select eligible star forming gas cells/particles. We use a maximum temperature
of $15000\;\mathrm{K}$, and a density threshold dependent on the resolution of
the different initial conditions we simulate. (see Section 3.1)

\subsubsection{Blastwave Feedback (Gasoline2)}
The blastwave feedback model was first described in \citet{Stinson2006}.
Blastwave feedback uses an analytic estimate of the maximum radius and lifetime
of a SNII blast wave derived from \citet{Chevalier1974} and \citet{McKee1977}
respectively. The energy of SN feedback ($E_{SN}=E_{51}10^{51}\erg$) is smoothed
over a radius of $R_E$, which depends on the ambient density $n_0$ and pressure
$P_0$:
\begin{equation}
    R_e = 10^{1.74}E_{51}^{0.32}n_0^{-0.16}(10^{-4}P_0/k_B)^{-0.20}\pc
\end{equation}
The gas that this energy is distributed in is subsequently prevented from
cooling (in order to prevent numerical overcooling). Cooling is disabled for a
timescale $t_{max}$ also set by the ambient density and pressure:
\begin{equation}
    t_{max} = 10^{0.85}E_{51}^{0.32}n_0^{0.34}(10^{-4}P_0/k_B)^{-0.7}\Myr
\end{equation}
The blastwave feedback model has been used for a number of
simulations, ranging from the McMaster Unbiased Galaxy Simulations
\citep{Stinson2010}, the Eris high-resolution Milky Way \citep{Guedes2011}, and
the recent Romulus large-volume simulation \citep{Tremmel2017}, where it was
also used to model feedback from AGN.

\subsubsection{MaGICC Feedback (Gasoline2)}
The MaGICC feedback model extends the \citep{Stinson2006} blastwave model by
including the effects of early stellar feedback from UV radiation. In addition
to the basic blastwave model, which handles energy input from SNII, 10\% of the
total UV flux from the stellar population contained within a star particle is
deposited into the surrounding 32 neighbouring gas particles as thermal energy.
This injects $\sim 2$ times the energy deposited by SNII.
Unlike with the SNII, these particles do not have their cooling disabled. This
model is designed to mimic the formation of {\sc Hii} regions and the
destruction of star-forming clouds by the radiation from massive stars. This
model has been used for the MaGICC simulations (\citealt{Stinson2013} describes
this model, and the simulations that are later used in MaGICC), as well as the
large NIHAO suite of zoom-in simulations \citep{Wang2015}.

\subsubsection{Superbubble Feedback (Gasoline2)}
Unlike MaGICC, the superbubble feedback model \citep{Keller2014} is not designed
to incorporate other forms of energy than SNII, but instead to better model the
evolution of SNII-heated gas. Superbubble feedback avoids overcooling by
depositing thermal energy and SN ejecta into resolution elements where it
evolves with a brief, two-phase state.  The cold, dense shell component of the
resolution element is evaporated by thermal conduction at a rate which depends
on the hot phase's temperature $T_{\rm hot}$, the conduction rate coefficient
$\kappa_0=6.1\times10^{-7}\erg^{-1}\K^{-7/2}\cm^{-1}$, and the SPH smoothing
length $h$. The rate is calculated following \citet{MacLow1988}:
\begin{equation}
    \frac{dM_{\rm hot}}{dt} = \frac{16\pi\mu}{25k_B}\kappa_0T_{\rm hot}^{5/2}h
\end{equation}
Once a particle has fully evaporated its cold component, it begins evaporating
cold neighbour particles using a similar rate.  \citet{Keller2015,Keller2016}
have shown how this model can drive strong outflows at high redshift, and
regulate the formation of stars in galaxies up to $M_{\rm vir}\sim10^{12}\Msun$.
The full details of this model can be found in \citet{Keller2014}. Importantly,
the superbubble model evaporates particles stochastically, using a RNG to
convert the evaporation rate into a probability analogous to the star formation
rate.  In general, the superbubble model is less sensitive to numerical
parameter choices such as star particle mass, timestep, etc. than the blastwave
feedback model, and results in much more realistic ISM evolution and star
formation histories \citep{Keller2015}.

\subsubsection{Delayed Cooling Feedback (RAMSES)}
The feedback model we have used in the simulations using RAMSES is the ``delayed
cooling'' model described in \citet{Teyssier2013}. This model injects feedback
energy into both the thermal energy of a gas particle as well as in a
non-thermal, exponentially decaying form in order to prevent overcooling in
unresolved regions. Rather than tracking separate energy and density variables,
as is done in the superbubble model, this model simply treats this non-thermal
energy as a passive scalar, which does not contribute to the pressure within a
resolution element (thus avoiding a double counting, since it is injected into
both the turbulent and thermal energy). The evolution of the thermal
$\epsilon_{\rm thermal}$ and turbulent $\epsilon_{\rm turb}$ energies are
governed by the following pair of differential equations:

\begin{equation}
    \rho \frac {\rm{D}\epsilon_{\rm turb}}{\rm{D}t} = \dot E_{\rm inj} -
    \frac{\rho\epsilon_{\rm turb}}{t_{\rm diss}}
\end{equation}
\begin{equation}
    \rho \frac {\rm{D}\epsilon_{\rm thermal}}{\rm{D}t} = \dot E_{\rm inj} -
    P_{\rm thermal}\nabla\cdot\mathbf{v}-n_H^2\Lambda
\end{equation}
If the non-thermal turbulent energy is large enough ($\sqrt{2\epsilon_{\rm turb}}
> 10\kms$), then cooling is disabled completely ($\Lambda = 0$). We used 
a dissipation time of $t_{\rm diss}=20\Myr$. \citet{Teyssier2013} note that this
model is qualitatively similar to the \citet{Stinson2006} blastwave model we
have used in our blastwave {\sc Gasoline2} runs.
\begin{figure}
    \includegraphics[width=\columnwidth]{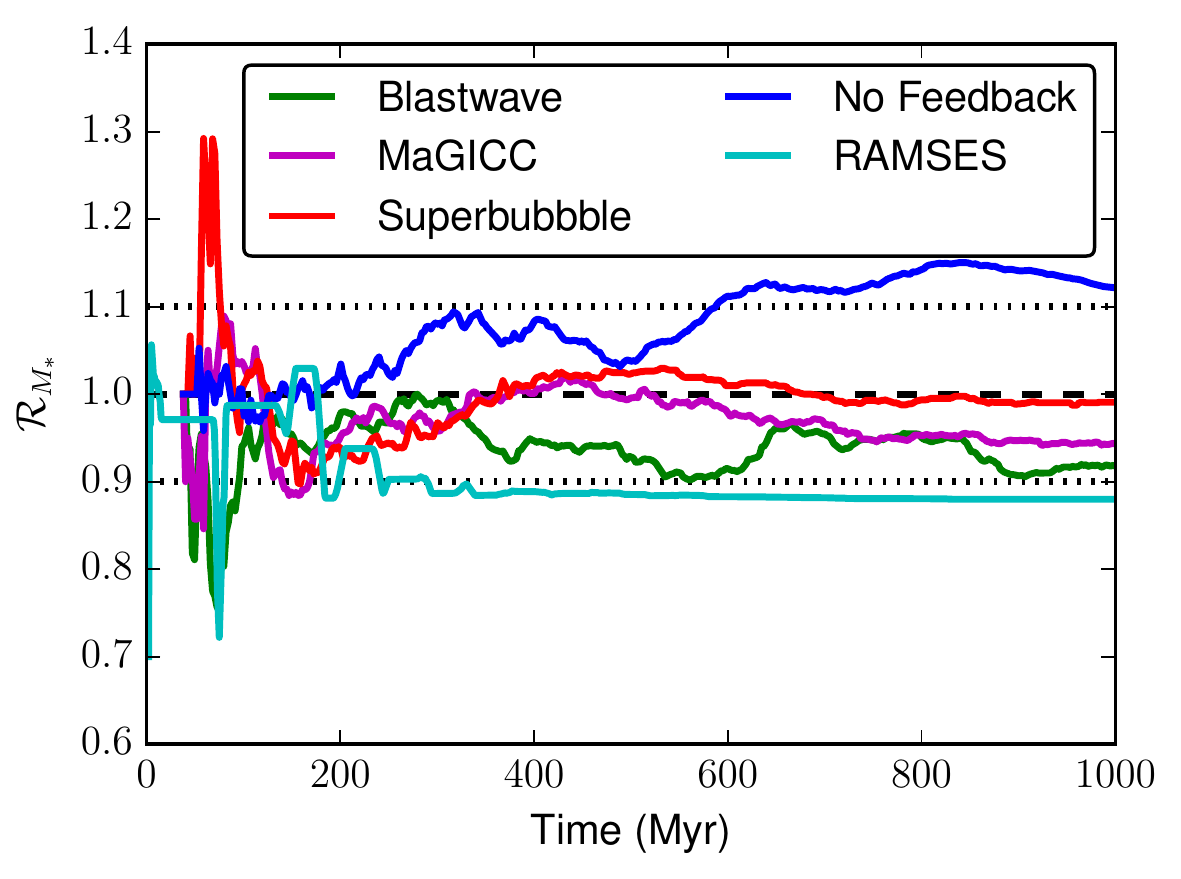}
    \caption{In this figure, we show the ratio of stellar masses formed in pairs
	of simulations run from identical initial conditions.  Each curve shows
	the ratio of the stellar mass formed in an isolated dwarf with different
	feedback models (Blastwave, MaGICC, Superbubble, and No Feedback) and
	different simulation codes ({\sc Gasoline2} for the former 4 curves,
	{\sc RAMSES} for the cyan curve).  If identical initial conditions and
	codes produced identical results, the ratio should be 1 (as shown by the
	black dashed curve).  Dotted black lines show variations in stellar mass
	of $\pm10\%$.}
    \label{dwarf_ratio}
\end{figure}
\begin{figure}
    \includegraphics[width=\columnwidth]{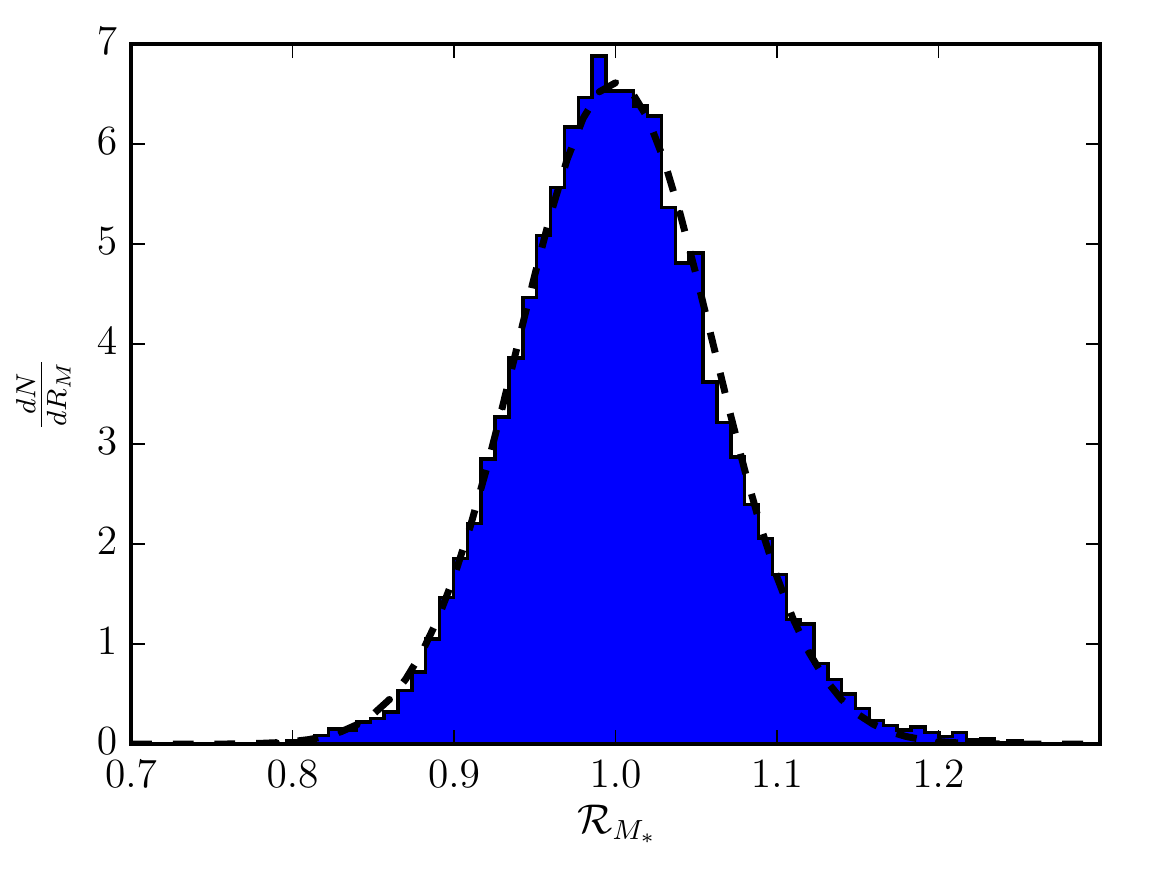}
    \caption{Histogram of the stellar mass ratios $\mathcal{R}_{M_*}$ at
	$t=3\Gyr$ for all pairs chosen from a set of 128 isolated dwarf
	simulations. Each of these galaxies began from identical initial
	conditions, and was evolved with identical code ({\sc Gasoline2} using
	superbubble feedback).  The solid histogram shows the distribution of
	mass ratios, with a fitted Gaussian shown by the black dashed curve. The
	mass ratios for these 8128 pairs are normally distributed, with
	$\sigma=0.06$ about $\mathcal{R}_{M_*} = 1$}
    \label{massratio_hist}
\end{figure}
\begin{figure}
    \includegraphics[width=\columnwidth]{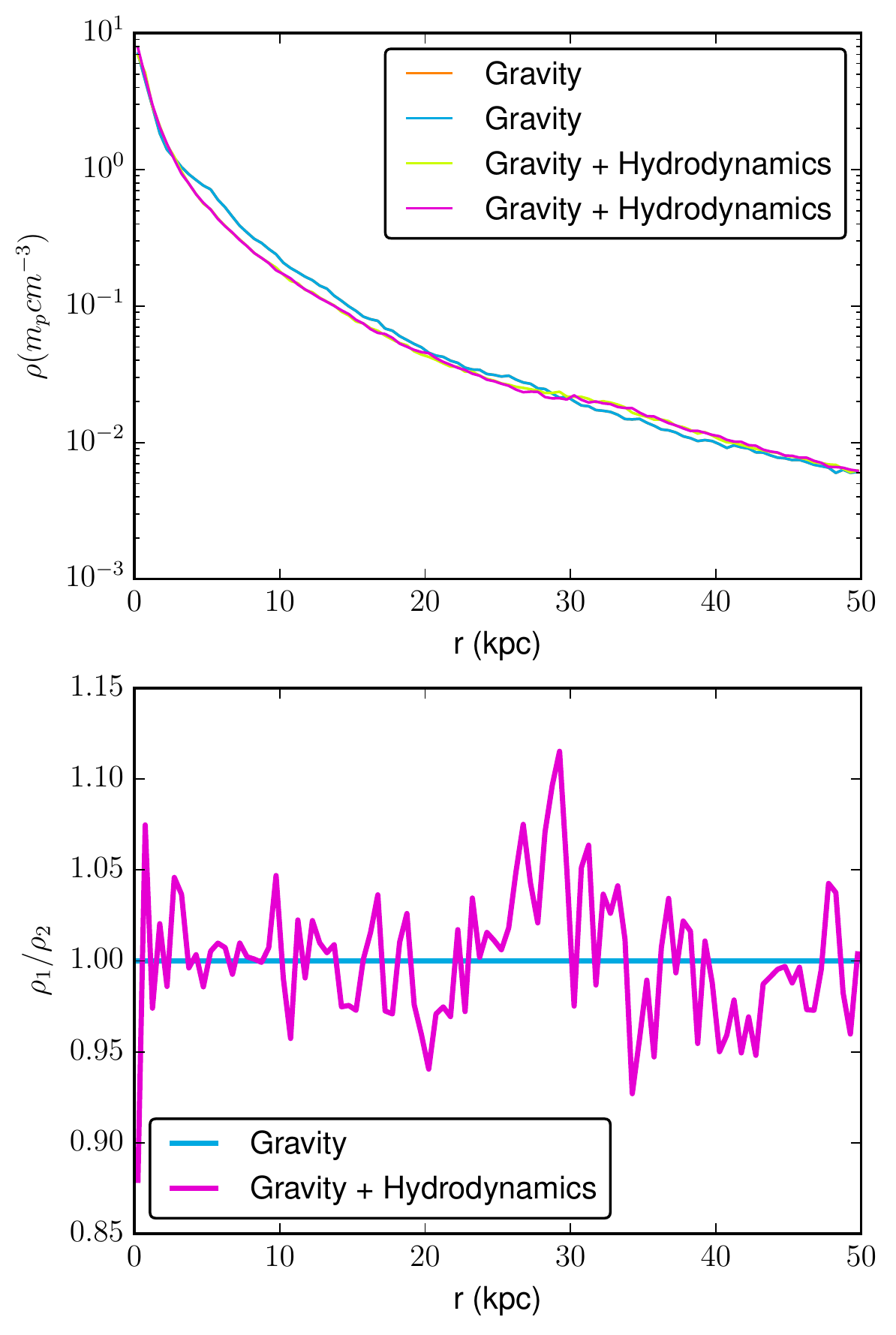}
    \caption{Without random number generators or parallel MPI reductions, as is
    the case when running {\sc Gasoline2} without hydrodynamics or star
    formation, identical initial conditions (run with the same number of
    processes) can produce bitwise identical outputs. Here we show our
    cosmological halo's density profile for two pairs of simulations, one run
    using gravity and hydrodynamics, and the other run using only gravity.
    The small perturbations between each run can be seen clearly in the bottom
    panel, where the gravity only case shows no difference run-to-run, while the
    case which includes hydrodynamics shows variations on the order of $10\%$
    from run to run.}
    \label{reduction_density_prof}
\end{figure}
\begin{figure}
    \includegraphics[width=\columnwidth]{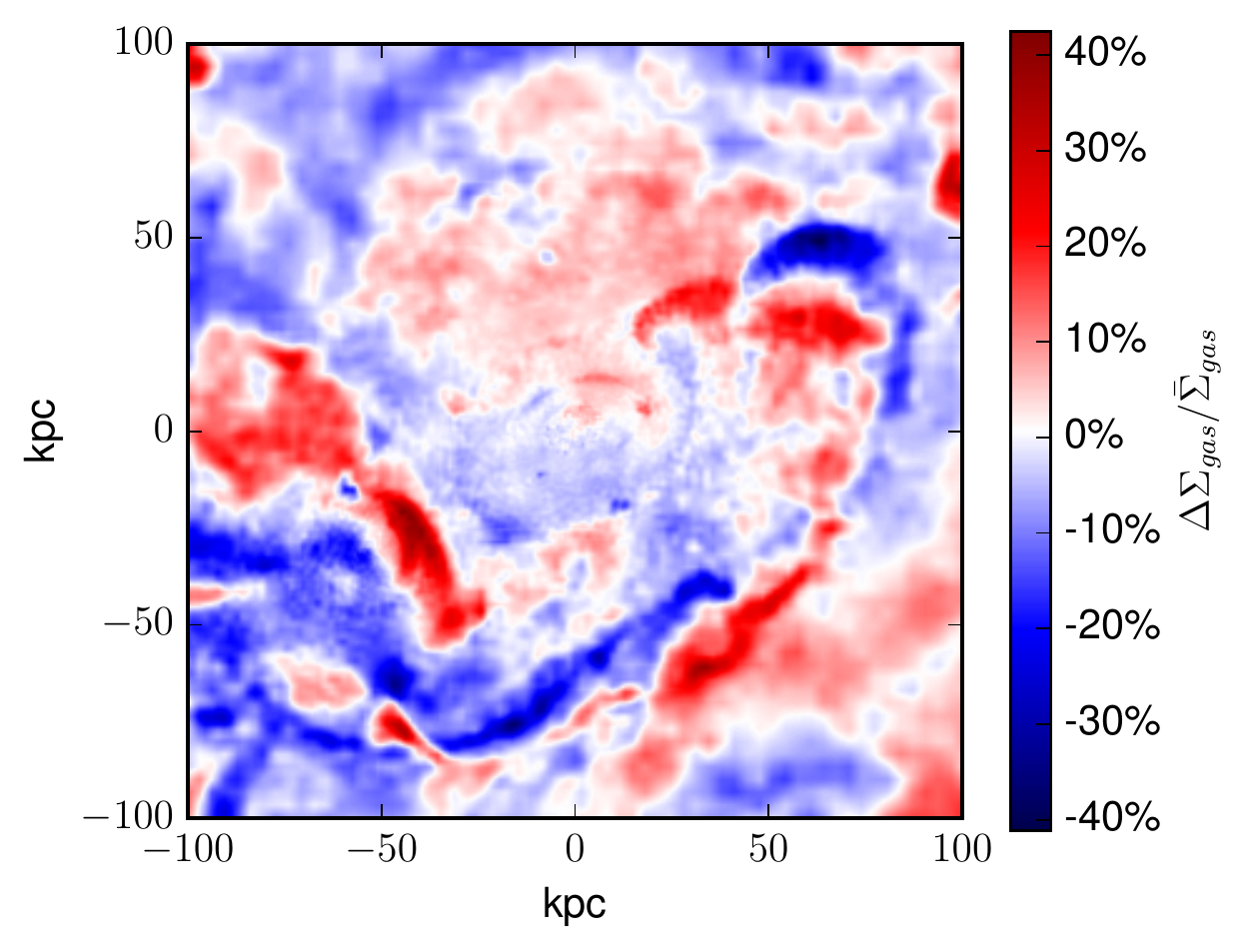}
    \caption{Here we show the difference in the column density of two
	pairs of cosmologically-evolved halos at $z=0.82$. Each of these four
	simulations are evolved from the same ICs, with identical
	code.  For one pair, gravity alone is used to evolve the IC, while the
	other uses gravity and SPH hydrodynamics.  In {\sc Gasoline2}, gravity
	calculations do not use parallel {\sc MPI} reductions, while SPH
	calculations do.  These reductions introduce roundoff-scale variations
	between the two runs with hydrodynamics, resulting in noticeable
	differences in the column density map.}
    \label{adiabatic_column}
\end{figure}
\begin{figure*}
    \includegraphics[width=\textwidth]{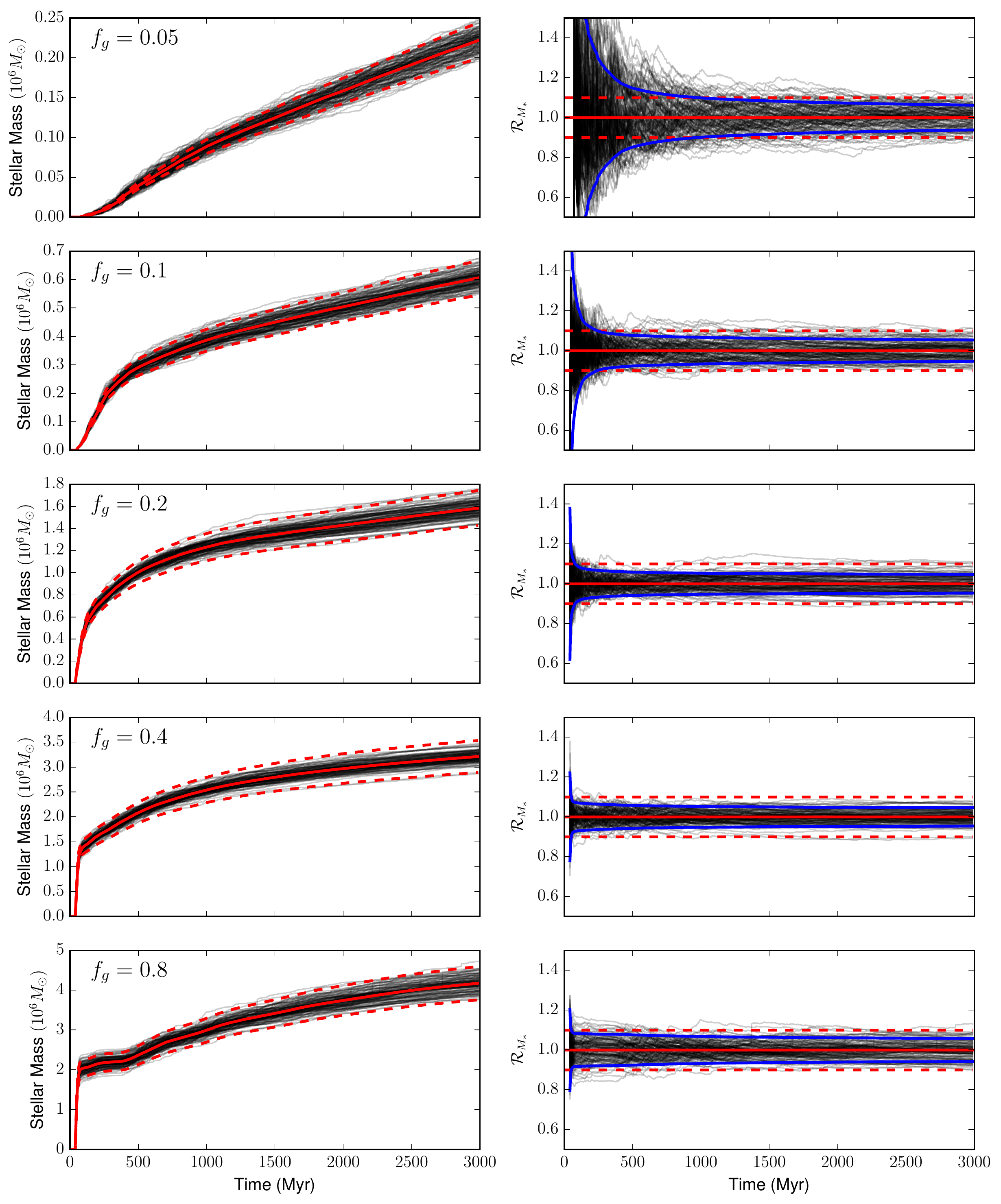}
    \caption{Even when varying the gas fraction $f_g$ of the disc in our 128
	dwarf simulations by a factor of 16, we can see in the left panels that
	stochastic inter-run variations rarely deviate by more than $\pm10\%$
	(red dashed curves) from the mean stellar mass (solid red curve). In the
	right hand panels, we can see that the variation from the mean
	$\mathcal{R_{M_*}}$ is fairly well bounded by $\sim 2/\sqrt{N_*}$ shot
	noise, where $N_*$ is the mean number of stars formed by all 128 dwarf
    galaxies (blue curves).}
    \label{manydwarfs_stellarmass}
\end{figure*}
\begin{figure}
    \includegraphics[width=\columnwidth]{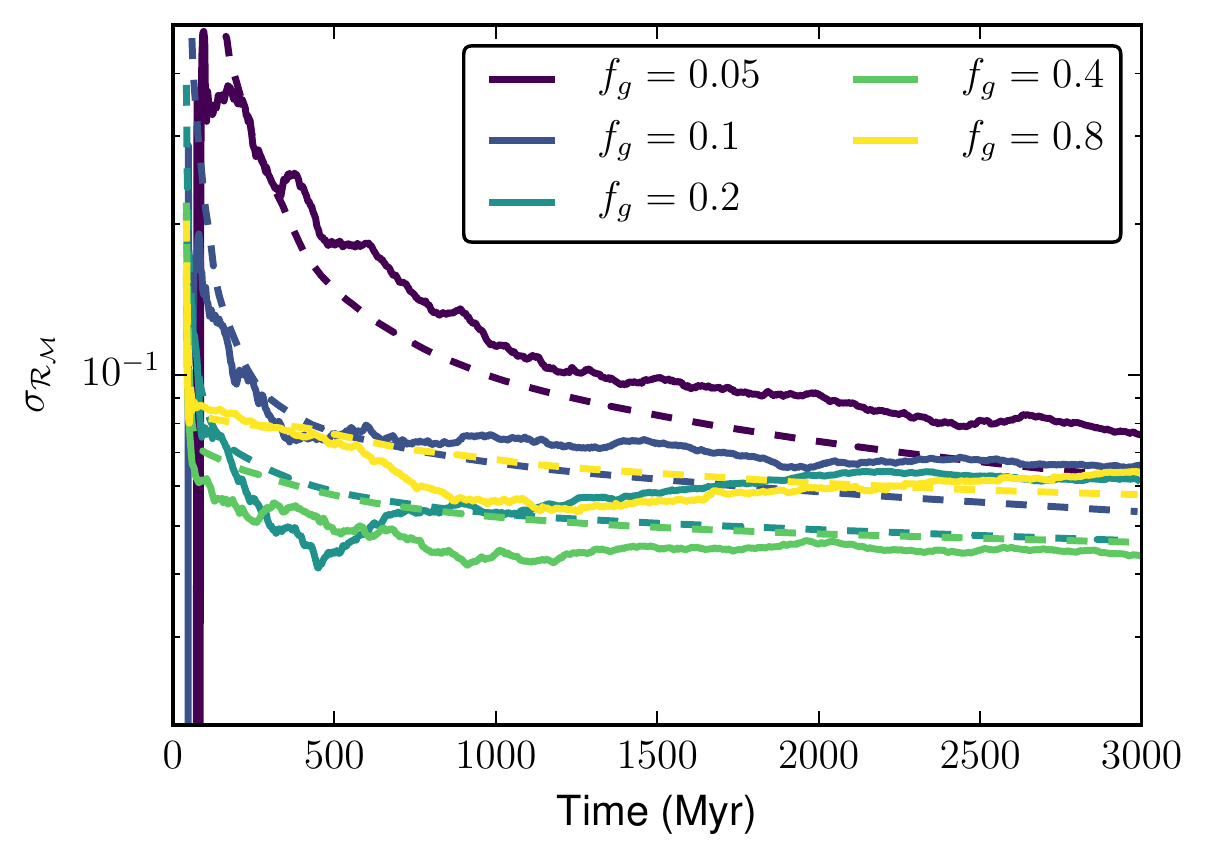}
    \caption{The deviation in stellar mass formed in each of our quiescent
	dwarves (solid curves), as can be seen here, is fairly well
	approximated by Poisson noise (dashed curves).  Stellar feedback is
	acting here to self-regulate star formation, and thereby reduce
	variations.}
    \label{manydwarfs_sigma}
\end{figure}
\begin{figure*}
    \includegraphics[width=\textwidth]{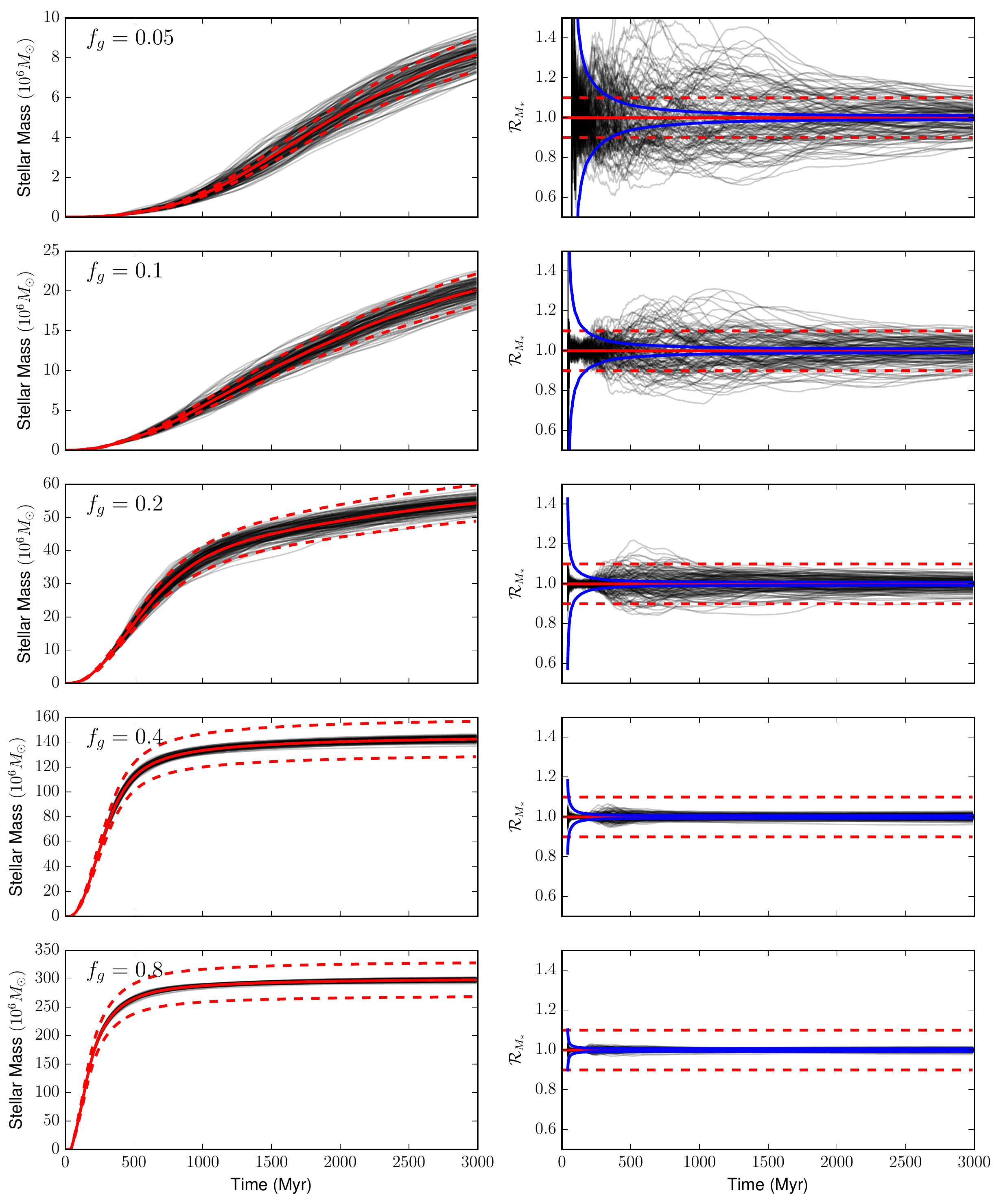}
    \caption{Unlike the case shown in Figure~\ref{manydwarfs_stellarmass}, when
	the same dwarf is evolved without any form of feedback, deviations in
	the star formation history can grow without self-regulation. These
	deviations cannot grow without bound, however, as can be seen in the
	higher gas fraction cases, where the deviations fall well below $10\%$.
	These cases show more uniformity due to simple gas exhaustion:
	in every galaxy, virtually all of the gas has formed stars.}
    \label{noFB_stellarmass}
\end{figure*}
\begin{figure}
    \includegraphics[width=\columnwidth]{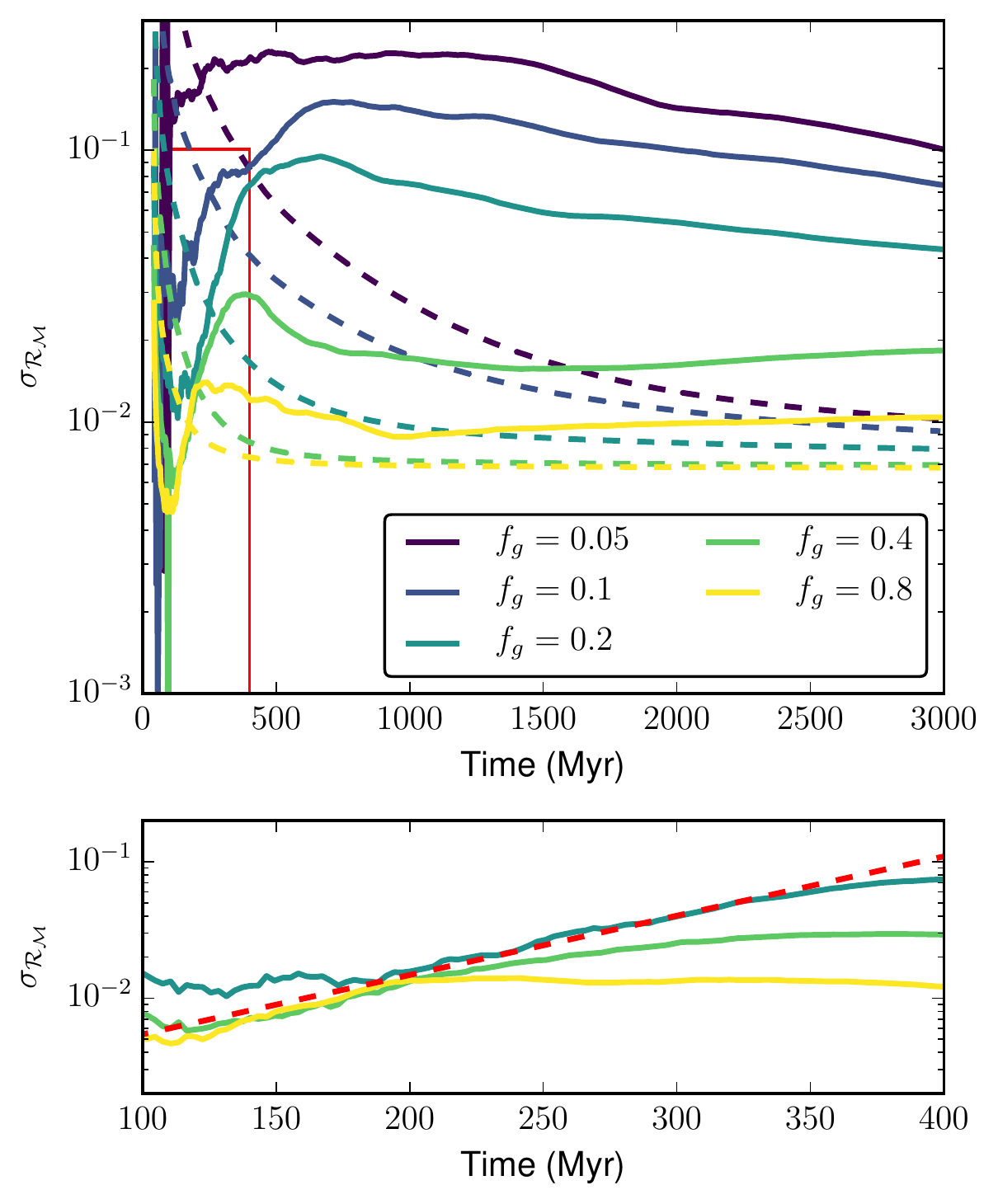}
    \caption{This figure shows the same evolution of stellar mass deviations
	shown in Figure~\ref{manydwarfs_sigma}, but without any form of stellar
	feedback. As one can see, while the scatter in stellar masses decreases
	at higher surface densities, it still far exceeds what we would expect
	from pure Poisson noise (dashed curves). Notice also that the peak in
	$\sigma_{\mathcal{R_M}}$ occurs at earlier times as we go to higher
	surface densities: gas exhaustion occurs earlier and earlier as the star
	formation rate increases. The bottom panel is a zoom of the region shown
	in the red rectangle. The dashed red line shows an exponential with an
	e-folding time of $100\Myr$.}
    \label{noFB_sigma}
\end{figure}
\begin{figure*}
    \includegraphics[width=\textwidth]{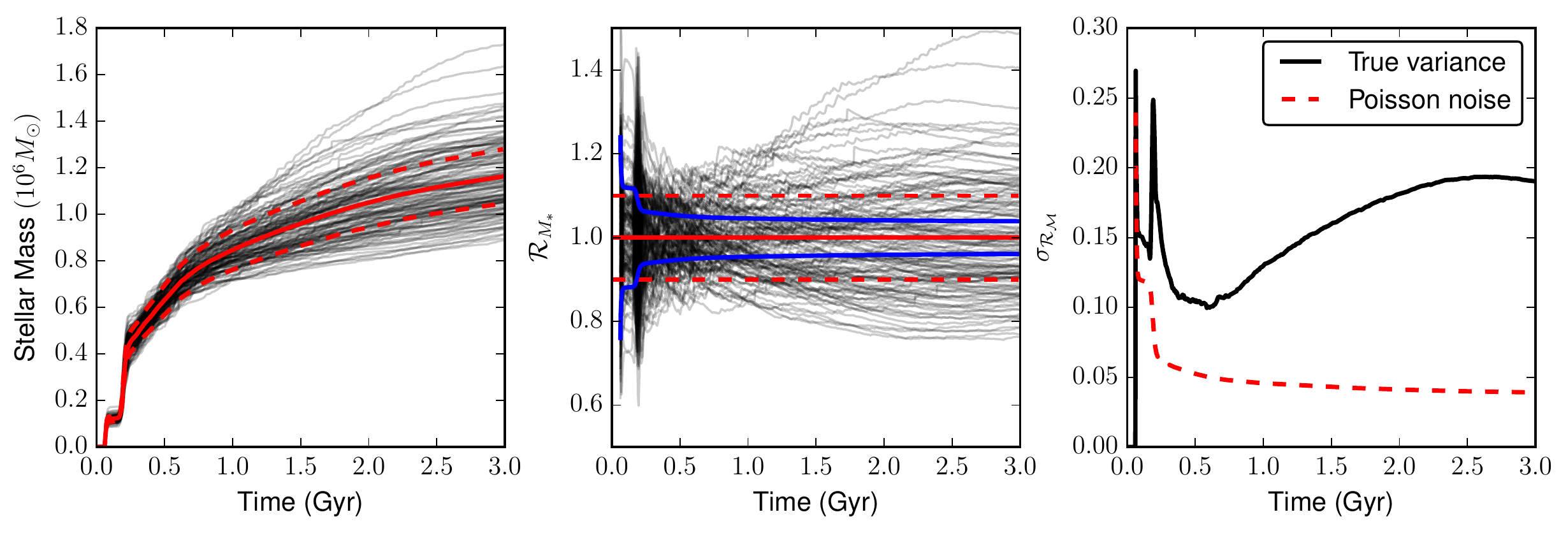}
    \caption{In a more violent system, such as this merger of two equal-mass
	dwarf galaxies, the starburst at $\sim200\Myr$ progresses differently in
	each system. Here we show the stellar masses formed (leftmost panel),
	the deviation in stellar mass compared to the mean (middle panel) and
	the evolution of the variance (rightmost panel) for the sample of
	mergers. The variation in stellar mass formed in this event is
	rapidly suppressed by feedback and gas exhaustion.  This variation
	leaves an indelible mark on each galaxy though, and after a few hundred
	Myr, the inter-run variation begins to grow as the star formation
	histories diverge from each other. As one can see in the left-most
	panel, once the merger occurs, the inter-run variation in stellar mass
	greatly exceeds the shot noise.}
    \label{mergers_stellarmass}
\end{figure*}
\begin{figure}
    \includegraphics[width=\columnwidth]{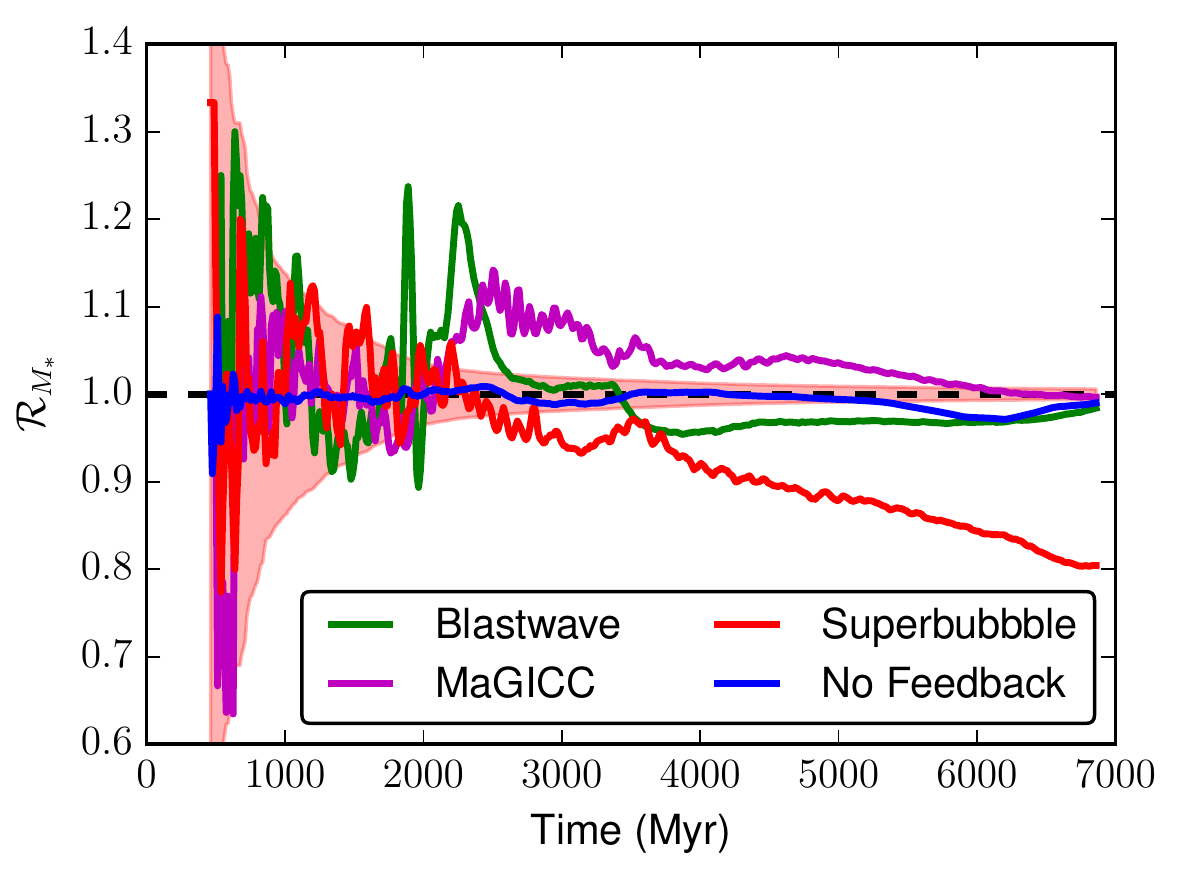}
    \caption{Here we show the same plot of deviations in stellar mass between
	pairs as was shown in Figure~\ref{dwarf_ratio}, but for a cosmological
	simulation of a Milky Way like galaxy. As the above figure shows,
	significant deviations in the stellar mass formed between two seemingly
	identical runs can grow and persist over Gyr timescales. The shaded
	region shows the $2/\sqrt{N_*}$ Poisson noise expected for the
	superbubble case. The dashed line at $\mathcal{R_{M_*}} = 1$ shows where
	the pair of runs form an equal number of stars.}
	\label{cosmo_ratio}
\end{figure}
\begin{figure*}
    \includegraphics[width=\textwidth]{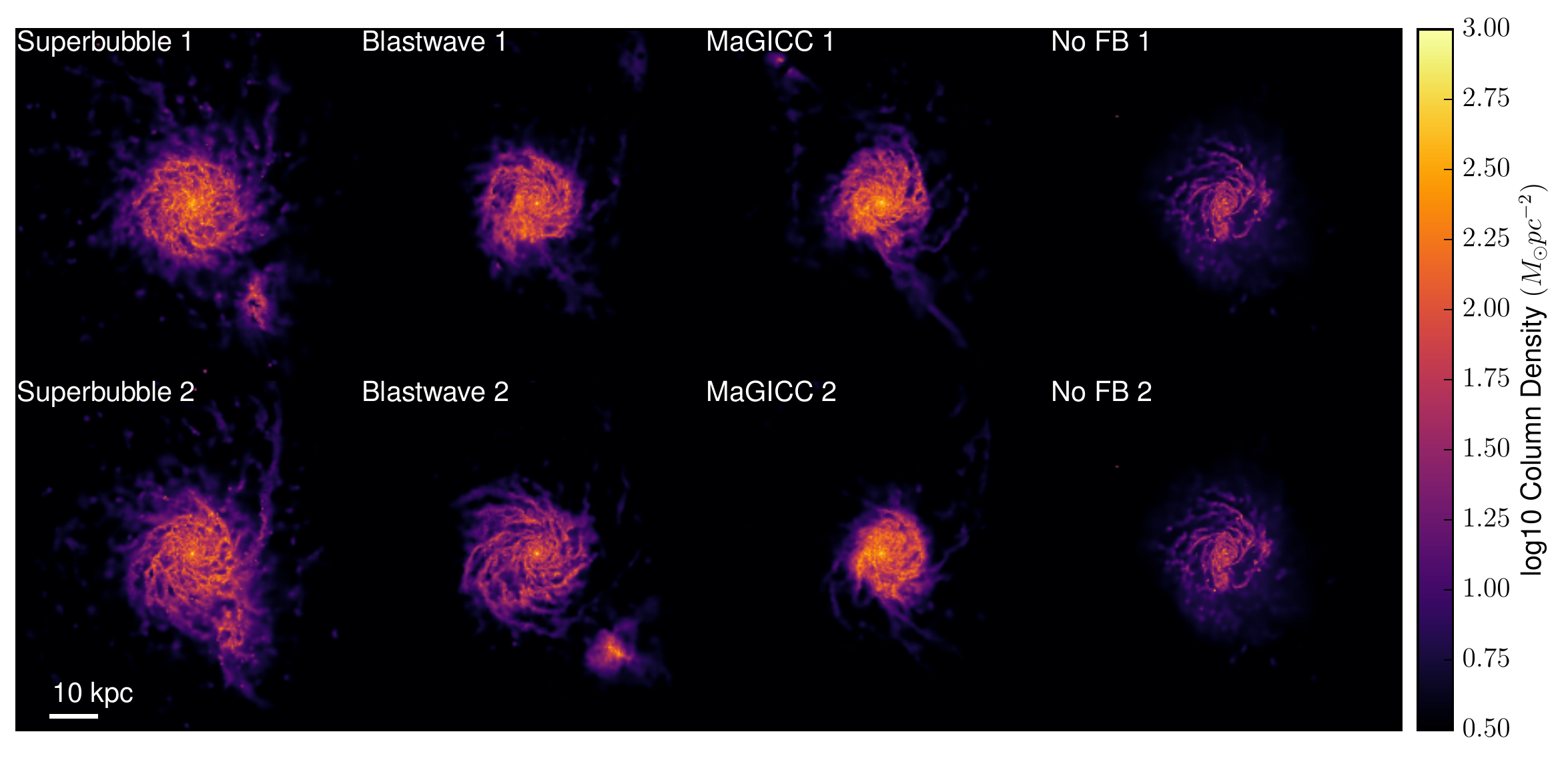}
    \caption{Gas column density from the four cosmological simulations. Each
    column shows a pair of simulations run with identical code and parameters,
    for four different feedback models. As is clear, significant variation in
    the timing of a recent merger has occurred, with some cases (Blastwave 1,
    Superbubble 2) having merged prior to their sibling runs.}
    \label{cosmo_column}
\end{figure*}
\begin{figure*}
    \includegraphics[width=\textwidth]{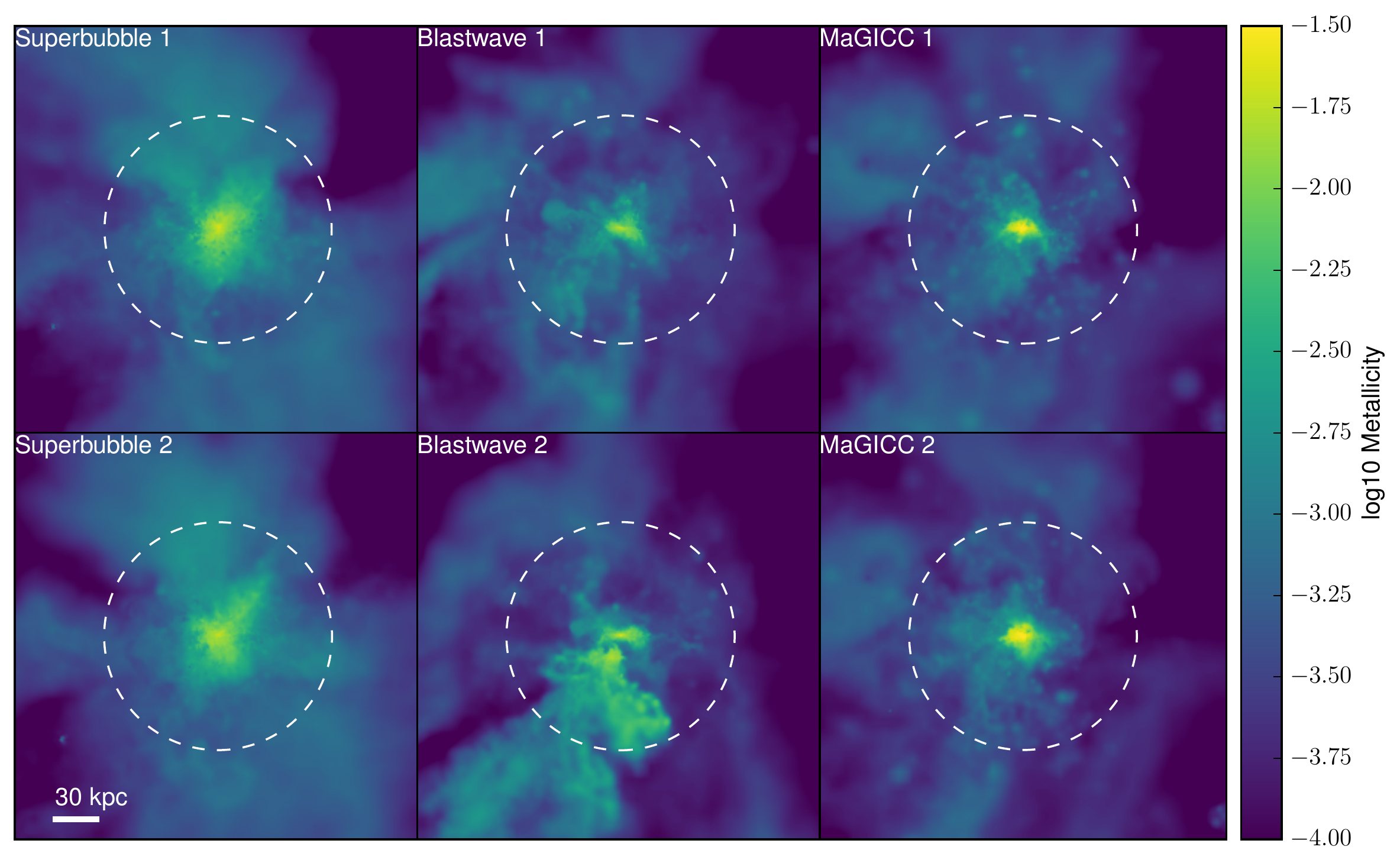}
    \caption{Column-averaged metallicity $\Sigma_{metals}/\Sigma$ for each of
    the three pairs of cosmological simulations that included feedback. Once
    again, there are clear variations in the amount and location of CGM metals
    within the virial radius (indicated by the dashed white circle).  The large
    difference visible in the blastwave case is due to the delay offset of the merger
    visible in the previous figure.  In case 2, the dwarf galaxy is polluting
    the CGM with metal-enriched outflows.  Meanwhile, case 1 has already seen
    this dwarf merge with the central galaxy, and the CGM has become more
    well-mixed.}
    \label{cosmo_metals}
\end{figure*}
\begin{figure}
    \includegraphics[width=\columnwidth]{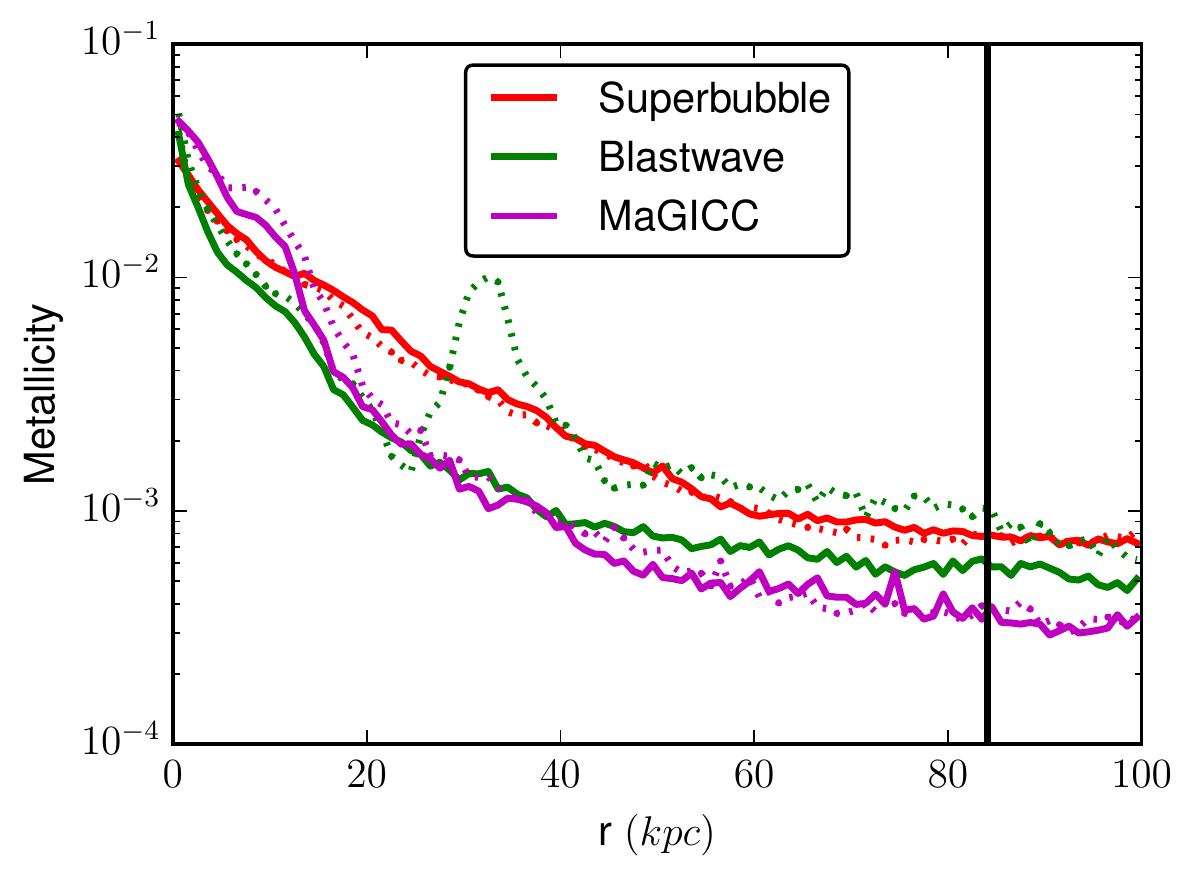}
    \caption{Here we show a mass-weighted radial profile of the metallicity in the
	halos of each of our cosmological galaxies.  Pairs using identical
	feedback physics are shown with identical colours.  The metallicity of
	the CGM is clearly significantly affected by the inter-run
	stochasticity. In fact, the change in merger timing between the two
	Blastwave feedback runs produced a change on the order of the difference
	between the three different feedback models. An unlucky choice of runs
	for a comparison could come to exactly the opposite set of conclusions
	than another set of runs, at least where it comes to the flow of metals
	in the CGM. The black line indicates the virial radius at $z=0.82$.}
    \label{cosmo_metal_prof}
\end{figure}
\begin{figure}
    \includegraphics[width=\columnwidth]{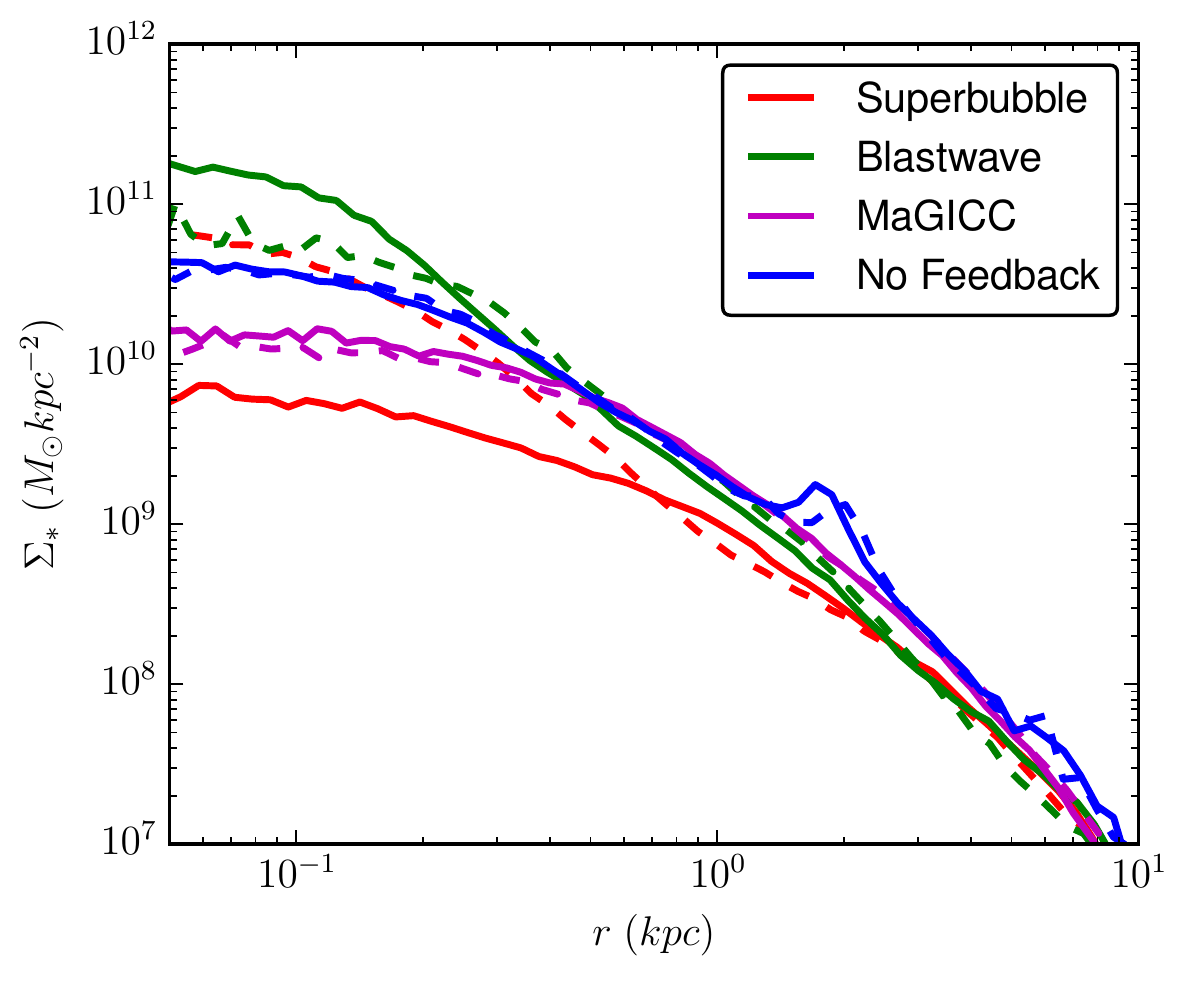}
    \caption{In these cosmological simulations, the stellar disc surface density
	profile also exhibits strong stochastic effects. Here, one can see that
	within $\sim400\pc$, the two Superbubble cases have nearly an order of
	magnitude difference in stellar surface density. Measurements of the
	stellar bulge in these two ``identical'' systems would find wildly
	different bulge masses. A random choice of two pairs of simulations to
	compare the effects of different feedback models could, once again, come
	to opposite conclusions based purely on the stochastic inter-run
	variation.}
    \label{cosmo_stellar_prof}
\end{figure}
\begin{figure}
    \includegraphics[width=\columnwidth]{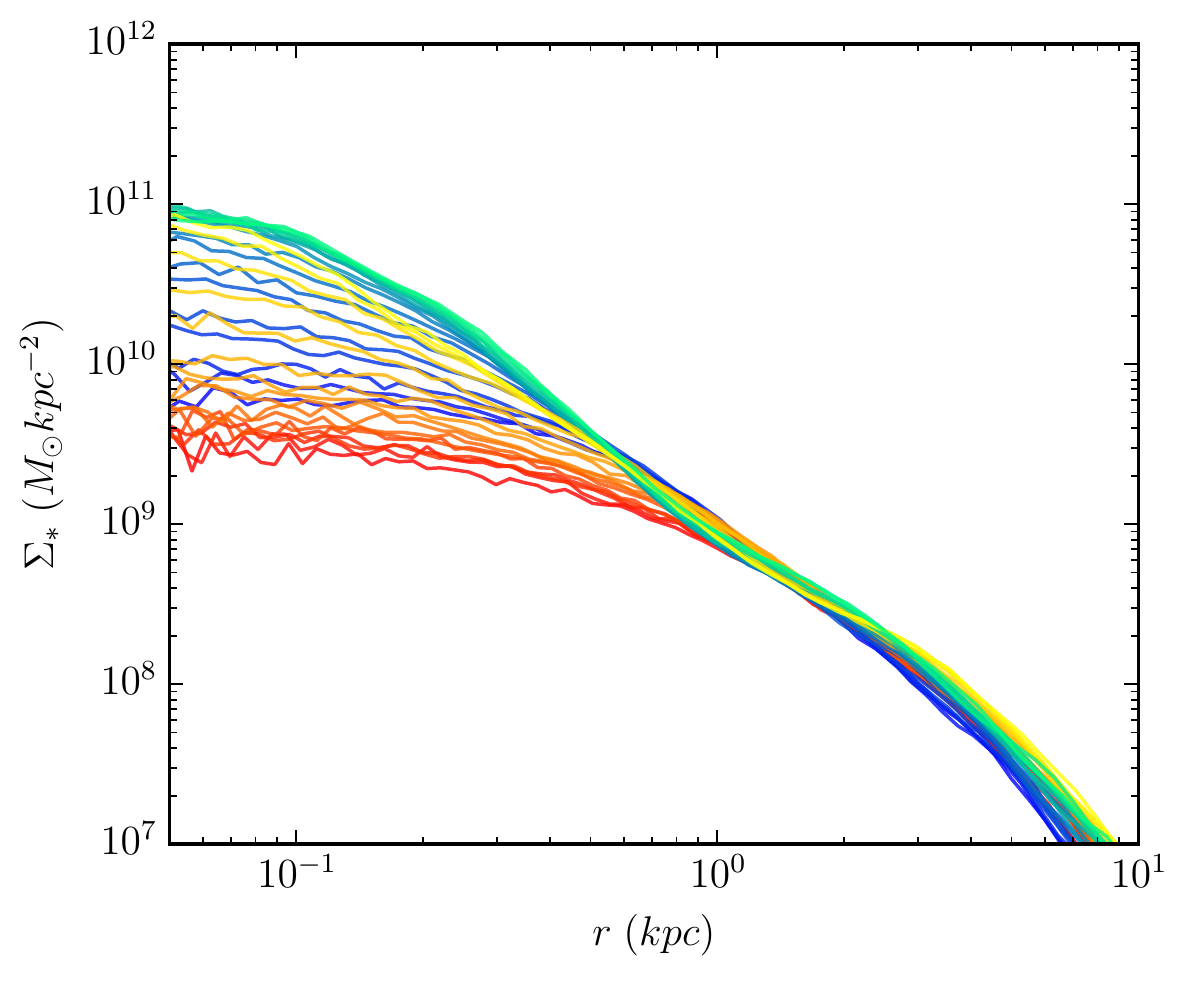}
    \caption{The run-to-run differences seen in the stellar density profile are
	often of similar magnitude to the deviations over a period of $\sim
	1\Gyr$ in individual simulations. Here we show the stellar surface
	density profiles for 5 snapshots before and 5 snapshots after the
	profiles shown in Figure~\ref{cosmo_stellar_prof}.  The snapshots are
	separated by $\sim100\Myr$.  We plot one of the superbubble runs using a
	blue-green gradient colourmap, and the other superbubble runs with a
	red-yellow gradient colourmap.  As you can see, these two sets of
	simulations cover roughly the same range over this $1\Gyr$ range.  As
	much of the variation occurs from noise introduced in the timing of star
	formation events, mergers, and inflows, averaging over a period
	comparable to the dynamical time of the galaxy can often give an
	indication as to how much stochasticity one may see in a resimulation.}
    \label{cosmo_stellar_prof_evo}
\end{figure}
\begin{figure}
    \includegraphics[width=\columnwidth]{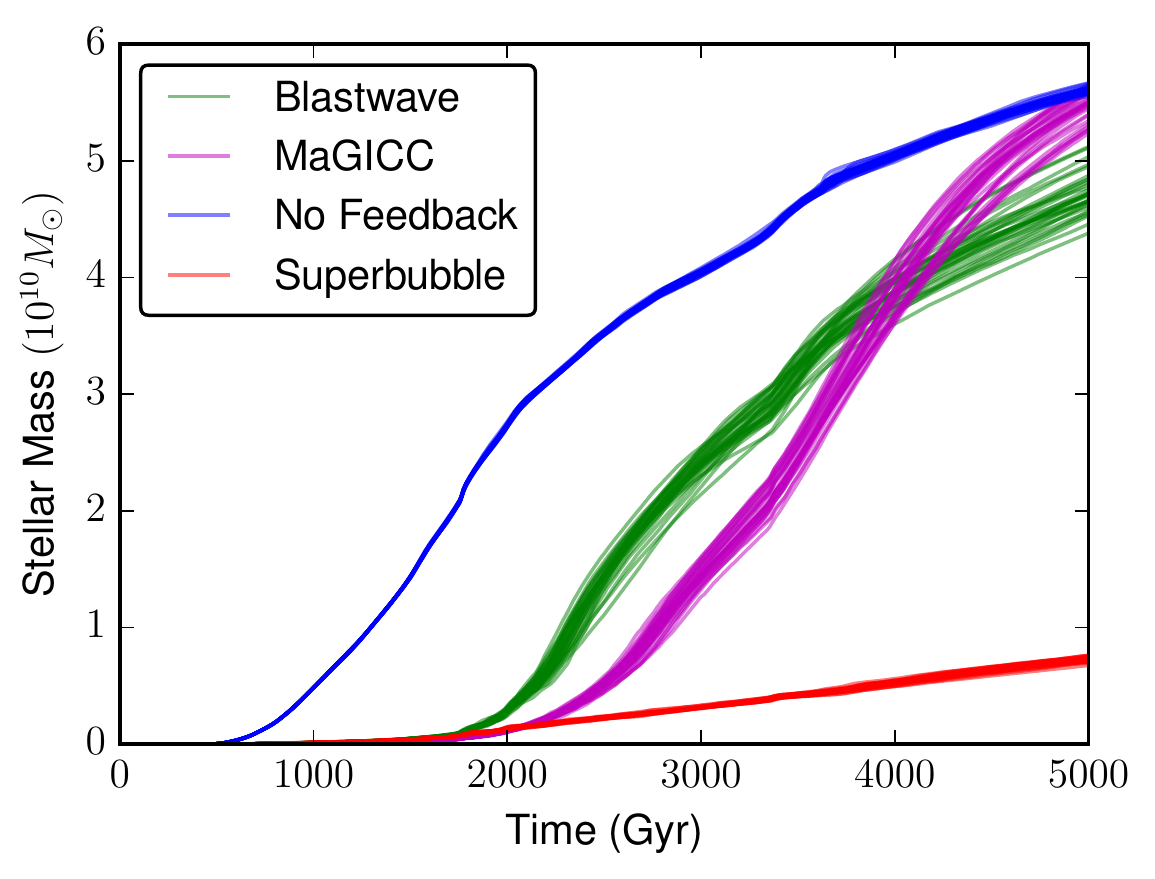}
    \caption{The stellar masses formed by $5\Gyr$ for each of the 32
    realizations of the same cosmologically-formed galaxy, using the three
    different feedback models shown earlier (as well as a case without
    feedback).  As is clear, the superbubble model forms significantly fewer
    stars than the two other feedback models.  Despite the scatter between
    different realizations, the difference between realizations with different
    feedback models is still distinguishable.}
	\label{cosmo_stellarmass}
\end{figure}
\begin{figure}
    \includegraphics[width=\columnwidth]{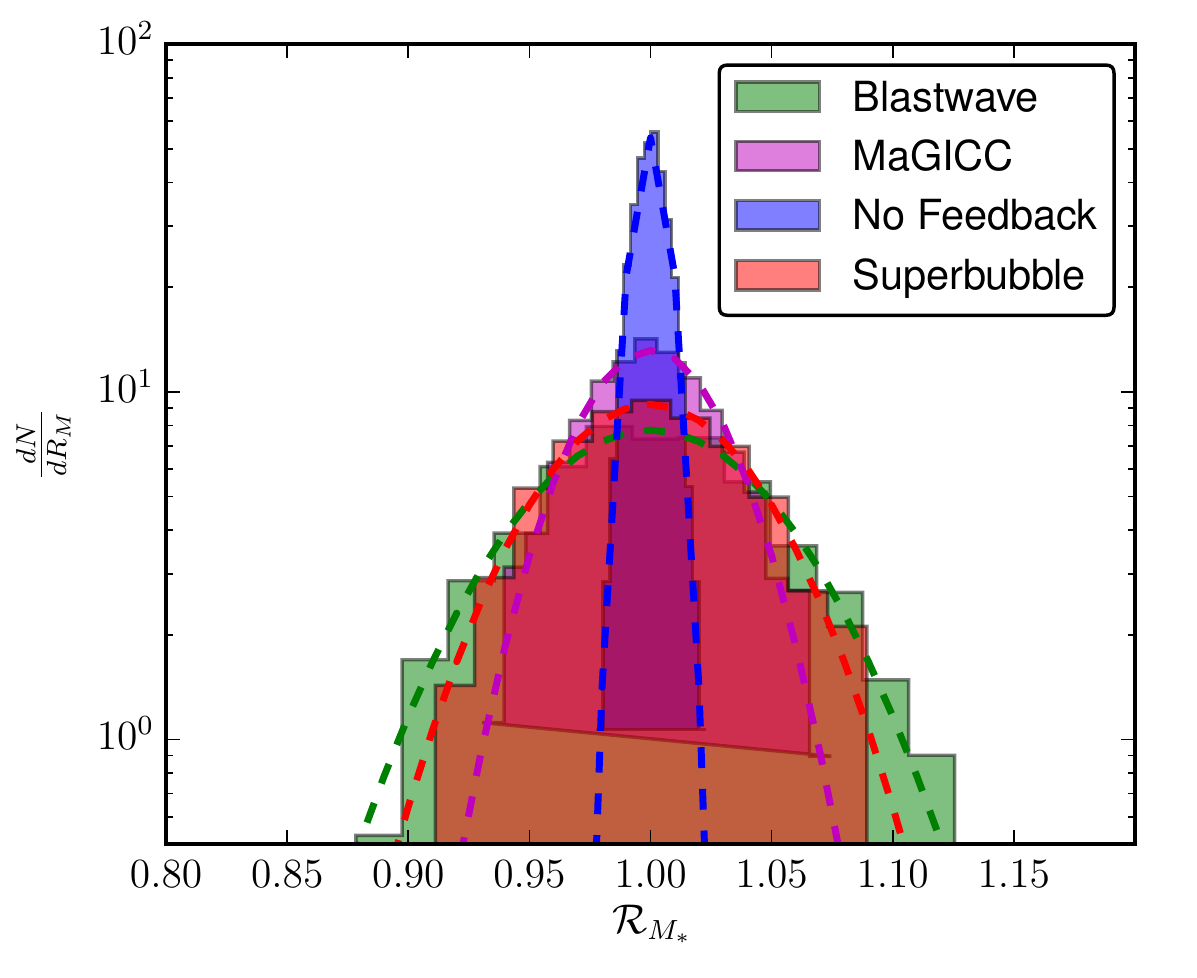}
    \caption{Histogram of the stellar mass ratios $\mathcal{R}_{M_*}$ at
    $t=5\Gyr$ for all pairs chosen from the 32 cosmological galaxy simulations.
    The solid histogram shows the distribution of mass ratios, with a fitted
    Gaussian shown by the dashed curve. As is clear, the realizations that
    include feedback, regardless of which subgrid model is used, have a much
    broader distribution of stellar mass than those without.}
    \label{cosmo_massratio_hist}
\end{figure}
\begin{figure}
    \includegraphics[width=\columnwidth]{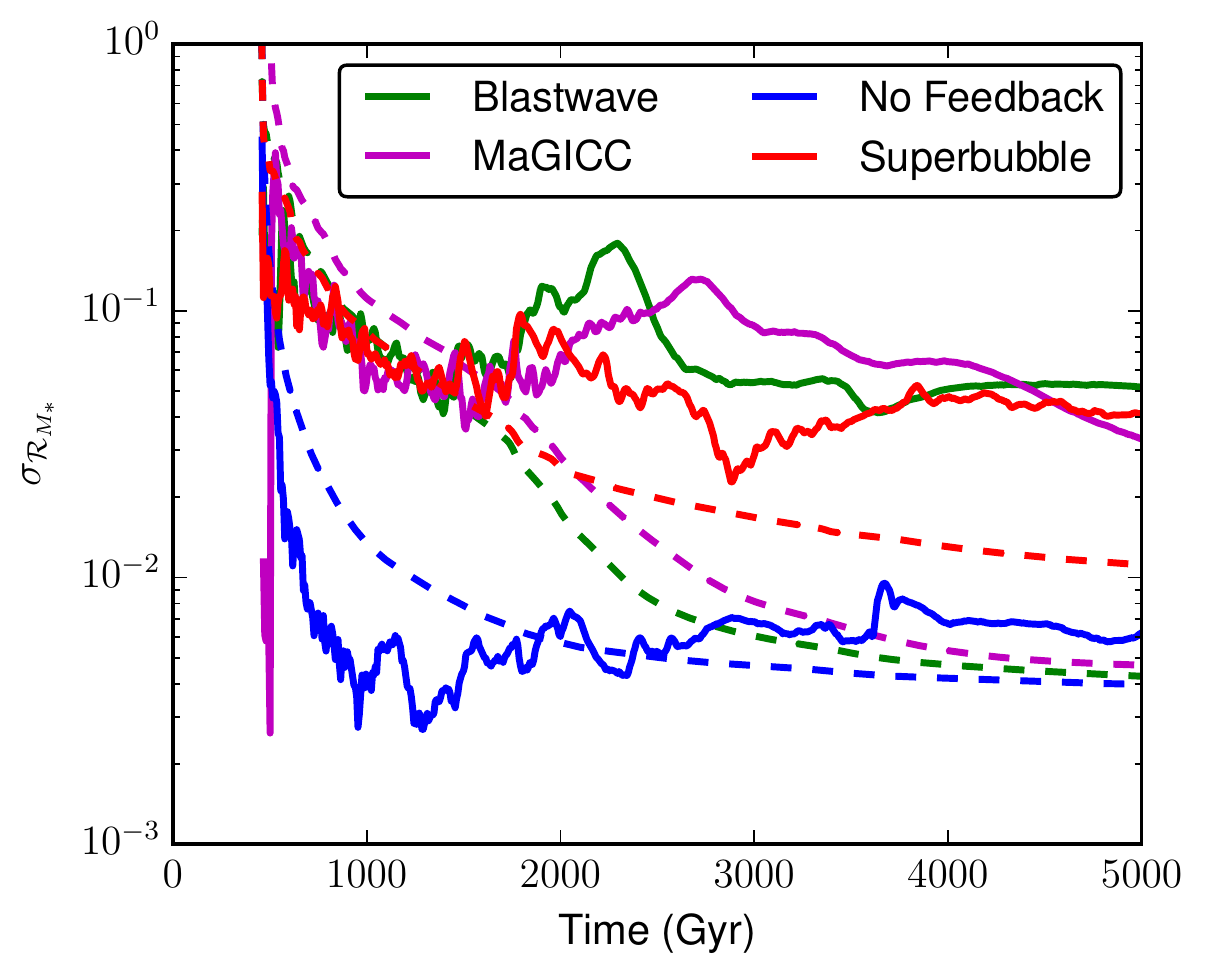}
    \caption{The time evolution of the scatter between the 32 cosmological
    realizations for each feedback model.  The $1\sigma$ scatter is shown with
    solid curves, while the $2/\sqrt{N_*}$ Poisson noise is shown by the dashed
    curves.  Despite having noticeably different star formation histories and
    final stellar masses, the three cases with feedback all have $\sigma\sim
    0.05$, roughly an order of magnitude larger than the scatter expected from
    Poisson noise alone.  Meanwhile, the scatter in the case without feedback is
    reasonably well approximated by a the Poisson noise due to discrete star
    formation events.}
    \label{cosmo_sigma}
\end{figure}
\section{Numerical Experiments}
\subsection{Does Stochasticity depend on code or subgrid model?} We begin by
demonstrating that run-to-run stochasticity is a universal behaviour,
independent of subgrid physics or simulation code. Using our isolated dwarf
initial conditions, we ran 5 pairs of simulations (2 simulations with blastwave
feedback, 2 with RAMSES, etc.). Each pair was run on the same machine with
identical ICs, parameters, compiled executable, and number of processors. We
calculate the ratio of stellar masses for each pair of simulations:
\begin{equation}
    \mathcal{R_{M_*}}(t) = M_{*,i}(t)/M_{*,j}(t)
\end{equation}

As one can see in Figure~\ref{dwarf_ratio}, each pair of simulations shows
variation on the order of $\sim10\%$ over the $1\Gyr$ we evolve them for. The
two cases with superbubble and MaGICC feedback appear to be closest to each
other. We next re-simulate the same IC, using superbubble feedback, 128 times.
We thus calculate $128!/(2!\cdot126!)=8192$ different values of
$\mathcal{R_{M_*}}$, one for each unique pair of simulations from the set of
128.  Again, all parameters and initial conditions are kept identical. As
Figure~\ref{massratio_hist} shows, the variation in stellar mass from run to run
is normally distributed, about $\mathcal{R_{M_*}} = 1$ (as we might expect),
with $\sigma_{\mathcal{R_M}}=0.06$. Thus, we can expect deviations of the order
$10\%$ from the mean stellar mass (at $1\Gyr$) in approximately $10\%$ of runs.

\subsection{Stochasticity without Random Numbers}
The simulations shown in Figure~\ref{dwarf_ratio} all include RNG-based star
formation models. As explained in section~\ref{floating_point}, parallel
floating-point reduction operations can result in non-deterministic behaviour.
In {\sc Gasoline2}, only the hydrodynamics is implemented using parallel
reductions for neighbour finding and force calculations\footnote{More precisely,
{\sc Gasoline2} uses a wrapper library, the Machine-Dependent Layer, {\sc MDL}
(\url{https://github.com/N-BodyShop/mdl}), for communication.  This library uses
asynchronous MPI communication to build a neighbour list during neighbour
finding.  While the neighbour list is guaranteed to always contain the same
members run-to-run, it does not guarantee that the neighbour list order is
identical.  This means that SPH force calculations and remote force summations
will have run-to-run roundoff-level differences due to the change in summation
order.}. This means that a pure N-body run of the same initial conditions should
be bitwise identical if run twice on the same number of threads. However, if the
same simulation is run with hydrodynamic forces included, small seed
perturbations in forces, on the order of $10^{-16}$, should be introduced in
each force calculation.  These seed perturbations may be amplified by the
chaotic behaviour of turbulence and self-gravitation.

In Figure~\ref{reduction_density_prof}, we show the results of a cosmological
simulation run without star formation or feedback, once with gravity alone, and
once with gravity and adiabatic hydrodynamics. This galaxy was evolved in each
case to half the age of the universe, $z\sim 0.8$. As is clear, the pure N-body
case produces exactly the same density profile for the largest halo in the
simulation, while the case with hydrodynamics shows noise in the density
profile.  The Median Absolute Deviation (MAD) in the density ratios
$\rho_{i,1}/\rho_{i,2}$ for all particles in the runs with hydrodynamics is 0.26
(we use the MAD here because of the long tail of large densities, which renders
less robust statistics inaccurate). If one examines the differences in the gas
distribution in these two adiabatic simulations, as
Figure~\ref{adiabatic_column} shows, deviations up to $>5\Msun \pc^{-2}$, or
$\pm40\%$ can be seen in the column density of the halo.  The MAD of the pixels
in this $500^2$ column density map is 5\%, or $0.17\Msun\pc^2$. The paired,
offset structures that are visible in the column density difference show one of
the major ways these perturbations grow in a cosmological simulation:
small-scale chaos saturates to change the timings of merger and inflow events.
Naturally, a change in the distribution and densities of gas within a galaxy
will result in different star formation activity, the formation of different
morphological structure, and may even change effectiveness of feedback by
altering the density of the ISM surrounding the sites of star formation.

\subsection{Stochasticity in different dwarf systems}
\begin{table*}
    \begin{tabular}{ccccccc}
        \hline
        $f_g$ & $M_{g,0}\; (10^6\Msun)$ & $t_{dep}$ (No Feedback)
        $(Gyr)$ & SFE &
        Maximum $\sigma_\mathcal{R_{M_*}}$ & Final $\sigma_\mathcal{R_{M_*}}$ &
        Maximum $\mathcal{R_{M_*}}$ \\
        \hline
        \hline
        0.05 & 21.5 & 252.1 (9.191)& 1.0\% (37\%) & 0.38 (0.20) & 0.08 (0.10) & 2.49 (2.79) \\
        0.10 & 43   & 78.48 (6.365)& 1.4\% (47\%) & 0.16 (0.15) & 0.07 (0.07) & 1.41 (1.78) \\
        0.20 & 86   & 29.85 (2.166)& 1.8\% (63\%) & 0.07 (0.10) & 0.06 (0.04) & 1.28 (1.44) \\
        0.40 & 172  & 21.11 (0.735)& 1.9\% (83\%) & 0.06 (0.03) & 0.04 (0.02) & 1.23 (1.12) \\
        0.80 & 344  & 24.62 (0.420)& 1.2\% (87\%) & 0.08 (0.01) & 0.06 (0.01) & 1.35 (1.06) \\
        \hline
    \end{tabular}
    \caption{Results from the sets of 128 dwarf simulations. Each row shows the
    initial conditions and final results from 128 identical ICs run with and
    without feedback. Values in brackets are those from the simulations without
    feedback. Maximum values are filtered to only include values after
    $100\Myr$ to remove the peak in $\mathcal{R_{M_*}}$ at the beginning of each
    run.}
    \label{manydwarfs}
\end{table*}
In Figure~\ref{manydwarfs_stellarmass}, we show the result of simulating our
dwarf with various gas surface densities, once again re-simulating each IC 128
times. We use the superbubble feedback model for these simulations. By varying
the disk gas fraction from $f_g = 0.05$ up to $f_g = 0.8$ (the original IC has
$f_g=0.2$), while keeping the total disk mass constant, we are able to vary the
overall star formation rate and gas depletion times. As can be seen from the
left hand panels, the galaxies at high gas fractions experience a strong initial
burst of star formation at $\sim 50\Myr$, and continue to form stars more slowly
for the remaining $3\Gyr$ of their evolution. Despite the wide difference in the
total number of stars formed across this range of gas fractions, the variation
in each of the five cases is typically of the order of the shot noise, as seen
in Figure~\ref{manydwarfs_sigma}. As that figure shows, there is a large
variation introduced by the initial burst of star formation, which rapidly
decreases as star formation continues. We here fit a Gaussian to the
distribution of $\mathcal{R_{M_*}}$ values, and plot the evolution of the
distribution's standard deviation $\sigma$ over time. Each of these cases has
standard deviation fairly well approximated as simple shot noise with $\sigma =
2N_*^{-1/2}$, where $N_*$ is the total number of star particles in the
simulation. Naturally, this implies that for these isolated, fairly quiescent
systems, the magnitude of this variation is resolution dependent, and will
converge towards zero at higher resolutions.  Of course, once individual stars
are resolved this variance will saturate out.

When the same set of simulations are run without feedback, we naturally would
expect to see more star formation (and lower shot noise). This is indeed what we
see in Figure~\ref{noFB_stellarmass}. As would also be expected, the higher gas
surface densities convert a larger fraction of their gas mass into stars.  The
right hand panels show that in the two highest gas surface density cases, the
standard deviation in $\mathcal{R_{M_*}}$ is well below $0.1$, as is the shot
noise.  Figure~\ref{noFB_sigma} shows this clearly: in each case, the final
shot-noise deviation is less than $0.02$. However, only the highest gas surface
density case shows an actual deviation this low. For every other gas fraction,
we see a peak in the stellar mass deviation many times larger than what we would
expect from Poisson noise alone. Before gas depletion becomes important, the
deviation grows roughly exponentially, with a Lyapunov exponent of $\lambda = 10
\Gyr^{-1}$, as the bottom panel of figure~\ref{noFB_sigma} shows. For the low
gas surface density case, the deviation exceeds $0.2$ for over a Gyr, and
remains as high as $0.1$ after $3\Gyr$ of evolution. Even without stellar feedback,
it is clear that the stochastic variations introduced by micro-scale chaos can
grow to significant differences in the stellar mass formed in these dwarf galaxy
systems. We can estimate the depletion time for each of these galaxies by using
the half-mass star formation time $t_{*,1/2}$ (to better estimate both the cases
where gas exhaustion has not yet set in and those which it has, we use the time
where half of the stellar mass is assembled $T_{*,1/2}$):
\begin{equation}
    t_{\rm dep} = \frac{M_{g}(0\Gyr)}{2*M_*(3\Gyr)}t_{*,1/2}
\end{equation}
The final star formation efficiencies ($M_*(3\Gyr)/M_{g}(0\Gyr)$) for each case,
along with the initial conditions and depletion times, are shown in
Table~\ref{manydwarfs}.  It is clear from figure~\ref{manydwarfs_sigma} and
figure~\ref{noFB_sigma} that stochastic variations in stellar mass grow
differently with and without stellar feedback.  Without stellar feedback, there
is a monotonic relationship between the gas depletion time, the variance in
stellar mass, and the Poisson noise expected.  When the depletion times become
short, the variance approaches the Poisson noise.  With feedback, however, the
relationship disappears, and instead we see that the variance approaches a
constant value of a few per cent, regardless of the depletion time.  As we shall
see in later in this paper, this appears to be the case in more complex,
cosmologically evolved galaxies as well.

\subsection{Stochasticity amplified by mergers}
In more extreme star formation events, such as a merger-induced starburst, the
run-to-run deviation can greatly exceed the $2N_*^{-1/2}$ value seen in the
previous runs. We built an initial condition by duplicating our $f_g=0.1$
isolated dwarf, and placing one copy $50\kpc$ at $45\deg$ from the galactic
plane of its twin (with zero initial relative velocity), giving us a simple
model of 1:1 major merger that occurs at $\sim200\Myr$.  When this merger
occurs, as can be seen in Figure~\ref{mergers_stellarmass}, a rapid starburst is
triggered. The variation in the timings of this starburst causes a brief spike
in the stellar mass deviation to $\sigma \sim 0.25$, along with a jump in the
total stellar mass (pushing the shot noise down). As the starburst proceeds and
eventually is extinguished by feedback, the stellar mass in the different runs
begin to converge again, with the deviation eventually falling to $\sigma=0.1$.
The variations in the timings and locations of star formation during this merger
is subsequently pumped up, returning the stellar mass deviation to
$\sigma\sim0.2$. Some pairs of dwarf merger remnants reach final stellar masses
after $2\Gyr$ that differ by more than a factor of two (the pair consisting of
the galaxy which forms $\sim1.5$ times the mean stellar mass and the galaxy
which forms $\sim0.7$ times the mean stellar mass).

\subsection{Stochasticity in cosmological galaxy formation}
Cosmological galaxy evolution, especially in the early stages at high redshift,
is characterized by many hierarchical merger events.
Figure~\ref{mergers_stellarmass} shows that numerical perturbations in mergers
can precipitate long-lived deviations in the star formation histories that
persist for Gyrs. Figure~\ref{cosmo_ratio} shows that this effect is important
in the more complex, realistic environment of a cosmological zoom-in simulation
as well.  We use {\sc Gasoline2} to simulate four pairs of the MUGS2 galaxy
g1536 to half the current age of the universe, $z\sim0.9$. Each pair is run
using the different stellar feedback models that were earlier compared in
Figure~\ref{dwarf_ratio}. As can be seen, deviations of greater than $10\%$ in
the total stellar mass formed between the two simulations (which once again, use
identical codes and initial conditions) occur and persist for Gyrs. These
deviations are much greater than the Poisson noise that our simple isolated
dwarf galaxy exhibited in Figure~\ref{manydwarfs_sigma}. The spikes seen in the
blastwave runs just before and after $2\Gyr$ are a result of timing variations
in a gas rich merger.

Even the qualitative appearance of these galaxies is noticeably different
between runs. Figure~\ref{cosmo_column} shows the face-on gas column density for
each of these cosmological galaxies. Each of the three pairs run with feedback
show noticeable differences in the progress of a merger. The superbubble and
blastwave cases each show one galaxy in the process of merging with a smaller
companion, while the other has not yet begun to merge. One of the MaGICC discs
has a clear tidal feature which is absent in its twin. Clearly, these
differences in the distribution of gas will cause differences in the
entire rest of the system.

The global star formation history is not the only large-scale property where
this stochasticity can be important. Stochastic differences manifest themselves
quite prominently in the distribution of CGM metals, as is shown in the metal
column densities of Figure~\ref{cosmo_metals}. As can be seen there, the merger
timing offset in the two blastwave simulations causes a noticeable plume of
metal-rich ejecta to be visible in one of the pairs, but absent in the other. If
we take a profile of the radial metallicity, as in
Figure~\ref{cosmo_metal_prof}, we can see that this difference is significant.
The two blastwave cases show a larger difference in the average metallicity from
$50\kpc$ out to the virial radius than any of the differences between any of the
other cases using different feedback physics.  When stochastic variation alone
can produce changes in measured quantities this large, it is imperative that
studies comparing different physical models take great care to show that the
differences between simulations are actually the result of the change in
physical model, and not mere stochastic variation.  A na{\"i}ve researcher
could, given the results shown in Figure~\ref{cosmo_metal_prof}, present
completely contradictory results depending on which of the two ``identical''
runs they happened to obtain for each of the simulations they compare.  It is
likely that the literature contains works that attempt to explain stochastic
differences between runs as the result of important physical differences.

This magnitude of variation can also be seen in the profile of the stellar disc,
as we show in Figure~\ref{cosmo_stellar_prof}. Once again, the variation between two
twin simulations, in this case the superbubble runs, show a larger difference in
the stellar surface density of the inner $400\pc$ with each other than they do
from the other simulations which use different feedback physics. All four cases
show a cored central distribution of stars within the inner $\kpc$, which is
likely due to the lack of resolved gravity with the softening length of
$320\pc$. This flat central region shows a difference in stellar surface
density between the two superbubble runs of roughly an order of magnitude.

For quantities like these, where strong run-to-run variations are present, it is
often the case that these are due to a fairly noisy, variable history. As we
show in Figure~\ref{cosmo_stellar_prof_evo}, the inner $\kpc$ of the galaxy
actually shows fluctuations of $\sim 1$ dex over a period of $\sim1\Gyr$. The
two superbubble runs shown there, one coloured yellow to red for different
times, each separated by $110\Myr$, the other coloured blue to green, vary from
themselves over this time range by roughly as much as they vary from each other.
The density of the central regions of the disc is clearly one of the properties
that variations in the timing, kinematics, and gas-richness of mergers can
strongly impact.  This could be expected, as mergers are known to be a powerful
mechanism for transporting angular momentum away from disc material, allowing it
to form a central bulge \citep{Toomre1977}.  The sensitivity of a given
quantity to the history of gas accretion and mergers on to a galaxy can often
be used as a measure of the stochastic variations you may see when comparing two
different runs.  By examining a window of time before and after critical
points like merger events, the uncertainty due to stochastic variation can
be estimated.

\subsubsection{Statistics of stochasticity in cosmological environments}
In order to better understand what is occuring in a fully cosmological context,
where galaxies are neither isolated but instead experience tidal interactions,
mergers, and continuous gas accretion, we have re-run the four sets of
simulations from the previous section 32 times each, for a total of 128
cosmological zoom-in simulations, using four different models for feedback.  We
run these simulations to $z\sim2$ ($5\Gyr$).

A concern these results raise is whether comparisons between, for example,
different feedback models may produce incorrect conclusions simply due to
the effect size being smaller than the intrinsic stochastic variation in the
quantity being measured for a galaxy.  We show here in
figure~\ref{cosmo_stellarmass} that, for comparing feedback processes, stellar
mass is a fairly robust indicator, with run-to-run scatter much less than the
differences between different feedback models.  Subtler effects, such as the
changes introduced by varying diffusion models presented in
\citep{Stinson2013,Su2017} do appear to produce changes in the stellar mass of
the same order as the scatter shown here.  In fact, the initial conditions
simulated here are exactly the same as those used in \citet{Stinson2013}, and
the results we see here suggest that there is in fact no statistically
significant difference between their ``Fiducial'' case and their case with ``Low
Diffusion''.

Figure~\ref{cosmo_massratio_hist} shows that the distribution of stellar masses
formed with each of the 4 feedback models are roughly Gaussian, like our simple
dwarf case shown in figure~\ref{massratio_hist}.  It is clear, however, that the
simulations run without feedback form a much narrower distribution of stellar
masses, while the three cases with feedback all have similar scatter, with
$\sigma\sim0.05$.  The variance of this distribution is, for the three cases
with feedback, constant to within a factor of a few, as is shown in
figure~\ref{cosmo_sigma}.  This is roughly an order of magnitude greater than
the variance expected from Poisson noise alone, and the three different feedback
models, despite producing quite different star formation histories, seem to all
share a roughly indistinguishable variance in the total stellar mass formed over
time.  Without feedback, however, we can see that the variance is significantly
lower, and in fact is comparable to the $2/\sqrt{N_*}$ shot noise over the
entire $5\Gyr$ history of these simulations. Without feedback, these galaxies
convert nearly their entire baryon budget into stars, suggesting that we are
seeing low variance in this case simply due to gas exhaustion.  The gas
depletion times in these galaxies are so short that any run which lags behind
the median has ample time to catch up, and those that rush ahead simply hit a
wall of mass conservation: they cannot form more mass in stars than they have in
baryonic mass.

\section{Discussion}
\subsection{Small effects are expensive to detect}
The clearest result of these experiments is that identifying and distinguishing
between subtle effects will require statistically meaningful samples and
analysis to confidently detect. Countless studies have looked at changes in the
star formation history, CGM/Outflow properties, and disc morphology effected by
different feedback models \citep[e.g.][]{Stinson2013,Smith2017,Nunez2017},
cooling schemes \citep[e.g.][]{Hu2016,Capelo2018}, hydrodynamics methods
\citep[e.g.][]{Hopkins2015,Valentini2017,Su2017}, and star formation recipes
\citep[e.g.][]{Su2018}. Each of these referenced papers presents
at least some results comparable in magnitude to what we show here arises
through stochastic variations alone. In order to be confident that these changes
are physical, numerical studies must present some analysis of the stochastic
variation intrinsic in their simulations.

\citet{Hoffmann2017} recommends 8 identical runs for each case to give two
samples per quartile of a uniform distribution. Our results suggest that
variations are normally distributed, which unfortunately means a large number of
cases are needed to sample the wings of the distribution. Despite that, for most
simulations of galaxies evolved over more than a Gyr, short-term deviations can
be used as an estimate for the uncertainty due to stochastic variations (for
example, see the peaks in the stellar mass ratio for each feedback-regulated
galaxy in Figure~\ref{cosmo_ratio}). In general, we find that feedback-regulated
systems (at least in the mass scale we have tested here) rarely deviate from one
another by more than $50\%$ in total stellar mass formed. Detecting the effect
of physical processes below this level will require either many (possibly dozens
of) identical resimulations, or the simulation of large systems (i.e.
cosmological volumes such as EAGLE, \citealt{Crain2015}, or Illustris
,\citealt{Vogelsberger2014b}, or large samples of zooms such as NIHAO
\citealt{Wang2015}, or E-MOSAICS, \citealt{Pfeffer2018,Kruijssen2018}).  Large
volumes and suites of zoom-in simulations provide a large enough sample of
galaxies, already perturbed by their different initial conditions, to look at
how different numerical models of physical processes manifest
population-averaged changes.  The variation within populations provided by these
larger simulation projects can be used to compare to the magnitude of effects
and evaluate their statistical significance.  Stochasticity and chaos are not
just the domain of dynamics: small-scale dynamical chaos couples to star
formation \& feedback, and can swamp smaller changes produced by the
experimenter.  Especially with systems involving mergers or bursty star
formation, comparisons must take these uncertainties into account.  While we do
not recommend every simulator take the extreme route of running hundreds of
copies of each simulation, a strategy similar to that advised by
\citet{Hoffmann2017} is almost certainly necessary for comparing small (factor
$<2$) variations in the quantities we have examined here. To the authors'
knowledge, only \citet{Su2017,Su2018} have re-simulated identical ICs to gauge
the uncertainty stochastic variations introduce in their simulations.
\citet{Su2017} uses a set of 5 runs, with ``re-shuffled'' feedback energy, while
\citet{Su2018} uses a pair of simulations with different RNG seeds.

Measuring the uncertainty due to stochastic variation can be done fairly simply
with re-simulations.  As Figure~\ref{massratio_hist} shows, stochastic
variations in quantities like the stellar mass of a galaxy tend to be normally
distributed.  A simple way to add confidence intervals to simulations would
be to select a reasonable point in the total lifetime of a galaxy (for
example, after the last major merger or the time at which the galaxy has
assembled half of it's mass), and re-simulate that galaxy from thereon a number
of times to generate a number of galaxy pairs.    For example, in a cosmological
simulation run to $z=0$, 16 pairs of galaxies run for $1\Gyr$ will yield 120
distinct pairs, and would require only slightly more computational time than
re-running the original simulation to $z=0$. A Gaussian fit to the ratios
between these pairs can then be used to determine the variance (which can then
be used to provide a $1/3/5\sigma$ confidence interval).  Even if a code does
not exhibit floating-point seeded stochastic differences, simply changing the
number of threads used, the least significant bits of parameters/initial
conditions, or the initial seed for the random-number generators can
be used to introduce the seeds for chaotic variation. In order to quantify the
amount of change that is due to the stochastic effects we examine here, you
would still need to run multiple simulations, regardless of whether a code
produces identical outputs when run without explicit perturbations.  The
direction of a the random walk, and the distance from the median value for a
given quantity is no better controlled in a reproducible code versus one which
introduces small, floating-point level perturbations. Differences between
simulations with alternative physical models which are smaller than this
confidence interval require multiple simulations to demonstrate that an
observed effect is not merely the result of stochastic variation.

\subsection{Self-regulation \& gas exhaustion are attractors}
One of the interesting results we have found is that not only do stochastic
variations exist for macroscopic quantities, but that the magnitude of these
variations can vary in different systems.  In those cases with depletion times
much less than the $3\Gyr$ we evolved our dwarfs for (the cases without feedback
and high gas surface densities), the final variation in stellar masses was
minuscule (with a $\sigma=0.01$ for the $f_g=0.8$ case).  Here the final stellar
masses have an upper limit due to mass conservation: a galaxy cannot form more
mass in stars than available in gas fuel.  They have a lower limit due to the
short depletion times relative to the time examined in the run: with an average
depletion time of $420\Myr$, even if one galaxy's star formation blazes ahead,
the others will have plenty of time to catch up in stellar mass once the early
star former exhausts its fuel. Without feedback, though, these galaxies wildly
overproduce stars, converting up to 87\% of their initial gas mass to stars in
the $3\Gyr$ we run these simulations for.  With more realistic,
cosmologically modelled galaxies, the violent environment in which the galaxy
forms drives nearly all galaxies towards star formation efficiencies of
$\sim100\%$ when feedback is omitted, with gas depletion times $<<t_{Hubble}$.
Pushing the gas depletion time of a simulated galaxy to very low values will, as
we have shown, reduce the final run-to-run variation in stellar mass, but at the
cost of no longer actually modelling real galaxies.  Observations reveal that
most galaxies have molecular gas depletion times $>\Gyr$ \citep{Bigiel2011}.

In the cases in which the gas depletion time is long, our
simulations comparing galaxies with and without feedback show two clearly
different outcomes.  Without feedback, stellar masses can diverge chaotically,
with a Lyapunov time of $\lambda^{-1}\sim 100\Myr$, as shown by the fit in
Figure~\ref{noFB_sigma}.  These variations can persist for a significant
fraction of a Hubble time.  With feedback, the behaviour is somewhat more
complex.  For both cosmological galaxies and quiescent, isolated systems, the
variance seems to be roughly constant, with $\sigma \sim0.05$.  There appears to
be no relation between the Poisson noise, depletion time, or even the feedback
mechanism, and the width of the stellar mass distribution for multiple runs.
Indeed, for our cosmological galaxies, which form many more stars (and thus have
much lower Poisson noise), we can see that feedback can actually {\it increase}
the variance compared to simulations run without feedback.

Stellar feedback acts to couple scales: star formation on small scales drives
the injection of energy and momentum on much larger scales.  As
figures~\ref{mergers_stellarmass} and~\ref{cosmo_sigma} show, small-scale
variations in the timing and progression of merger events can, through stellar
feedback, couple to larger scales and drive variations in the star formation
history of the galaxy that persist for $\Gyr$ or more.  However, these
deviations cannot grow without bound. For our cosmological, MW-like galaxy, each
of the three independent feedback models induced variance of $\sigma\sim0.05$
for the first $5\Gyr$ of the galaxies evolution, roughly an order of magnitude
above the Poisson noise.  We see in figure~\ref{cosmo_sigma} that a major merger
event at $\sim1.5\Gyr$ drives up the scatter in each case, producing a similar
effect as our idealized merger case shown in figure~\ref{mergers_stellarmass}.
Despite this, the variance never exceeds $\sigma\sim0.2$, despite sufficient gas
supply to more than double the stellar mass in both the case of our idealized
dwarf, and the cosmological galaxy (at least in the superbubble feedback case).
Together, all these results suggest we have at least two ``attractor''
solutions: a gas exhaustion/depletion equilibrium and a self-regulation
equilibrium.  In the case where depletion times are short, and feedback is
inefficient, simple mass conservation prevents galaxies from exceeding the
median mass, while the rapid conversion of stars to gas brings galaxies below
this mass quickly towards it.  When feedback is efficient, and thus depletion
times are longer, the coupling of star formation to large-scale outflows, disc
turbulence, etc. keeps the variance larger than expected from simple Poisson
noise, but prevents it from growing without bound.  A stochastic deviation that
decreases the star formation will allow more gas to collapse and form stars
(pushing the stellar mass up towards the attractor mass), while one that
increases it will disrupt and heat more gas via feedback, thus pushing the
stellar mass down towards this attractor solution.  While each realization will
evolve near this attractor solution, we can expect to find a not insignificant
scatter around it.

\subsection{Integrated quantities are less chaotic than instantaneous ones}
In general, quantities that vary over shorter timescales are more susceptible to
small-scale chaos. For example, while the feedback-regulated dwarf galaxy with
$f_g = 0.2$ has a maximum $\mathcal{R_{M_*}}$ of only 1.28, its SFR,
smoothed over $20\;\Myr$, shows maximum variations of more than a factor of 10
over the entire $3\Gyr$ lifetime. This too is a manifestation of the
self-regulation of star formation: when one galaxy begins to form stars at a
rate above the mean, feedback ensures that its star formation rapidly drops,
pushing it below the mean. The net result is that multiple runs show SFRs which
oscillate incoherently about a mean SFR, with phase offsets introduced by
small-scale stochastic variations.

Other quantities that can change on (relatively) short timescales will also see
large run-to-run variations. In particular, cosmological simulations can be
sensitive to the timings of merger events. A comparison made between
simulations that have recently experienced a major merger can be strongly
impacted by slight offsets in the timing or trajectory of that merger, as can be
seen in the CGM metallicity in Figures~\ref{cosmo_metals} and~\ref{cosmo_metal_prof}.
Unfortunately, as our toy merger shows, it may take a significant amount of time
for such differences to wash out. If a system experienced a major merger at
relatively low redshift, a $z=0$ comparison between two runs will likely be
strongly affected by variations in the merger timings that occur between the two
runs.

The variation in a single run over a short period of time can be used as a rough
estimate as to whether run-to-run variation may be large enough to confuse a
comparison. Looking at the instantaneous SFR in a short time window, for
example, will be a poor measure to use in a comparison of bursty systems, as
phase differences alone can produce large differences. As we showed in
figure~\ref{cosmo_stellar_prof_evo}, while the stellar profile of our disc
exhibits a factor of $\sim10$ variation in the peak surface density from run to
run, this stochastic variation is comparable to the variation in any single run
over $\sim1\Gyr$.

\subsection{Numerical considerations for reproducibility}
Recent work in the stellar \& planetary dynamics community has introduced some
novel techniques for minimizing numerical seeds of chaotic \& irreproducible
behaviour in simulations of N-body systems
\citep{Boekholt2015,Rein2017,Dehnen2017,Rein2018}.  As \citet{Laskar2009}
showed, errors on the order of a few hundred microns in the positions of a
planet can lead to wildly divergent outcomes of a planetary system, and thus
comparisons between different studies can be extremely difficult.

\citet{Rein2017} presented modifications to the collisional N-body simulation
package {\sc REBOUND} that allows for machine independent, bit-wise reproducible
simulations from identical initial conditions. {\sc REBOUND} is written in the
{\sc C99} standard (see section~\ref{floating_point} for why this is important),
and the authors have also implemented their own version of the {\tt pow()}
function. {\tt pow()}, {\tt exp()}, {\tt sin()}, and {\tt cos()} functions are
all implemented in hardware-dependent ways, meaning different CPUs from
different manufacturers, or even different CPU families from the same
manufacturer may produce different roundoff behaviour when these functions are
used. \citet{Rein2017} also implement a new binary output format that stores both the
initial conditions, the version of {\sc REBOUND} used to produce the simulation,
as well as the Jacobi coordinates used for the \citet{Wisdom1991} integrator,
rather than inertial coordinates, to prevent roundoff in recalculating these in a
resimulation. These features allow bitwise reproduction of solar system
simulations.

The same authors presented a different solution to the problem of reproducibility
in \citet{Rein2018}. This paper introduces a new integration scheme, {\sc
JANUS}, a high-order leap-frog integrator operating on 64-bit integers. As
integer arithmetic is commutative, associative, and distributive, this means that
{\sc JANUS} is bitwise identical regardless of compilation or parallel
communication details. As roundoff is symmetric in time, this allows {\sc
JANUS} to be reversed in time from a final state, returning to the same initial
conditions it began with. While this approach is well-suited for N-body solar
system models, it is unlikely it would be applicable to galaxy simulations,
which typically deal with much higher dynamic ranges, as well as additional
physical processes unsuited for integer-arithmetic integration. And of course,
as soon as non-adiabatic processes (radiative cooling, shocks, etc.) are included,
reversibility becomes impossible.  Beyond this, the value of ``reproducibility''
is questionable when the physical system itself is formally chaotic.

Developing a reproducible code, whether from scratch or by adapting an existing
code, can be a complex task.  A large amount of software infrastructure would
be required to support this: either algorithms would need to be re-written to
avoid reductions, or alternative implementations of the {\sc OpenMP} and {\sc
MPI} reduction functions would need to be written\footnote{This can be seen with
{\sc Gasoline2}'s custom interaction-list based parallelism for gravity.  The
interaction lists used for gravity are inherited from {\sc pkdgrav}, and forgoes
the {\sc MPI} reductions that are used for SPH sums in {\sc Gasoline2}.  This is
ultimately why the gravity-only {\sc Gasoline2} runs are reproducible, while
runs with hydrodynamics are not.}.  All sub-resolution models that rely on RNGs
would require careful accounting that all RNGs are seeded identically, and that
the RNG is identical across different machines and operating systems.  This too
may require a complete re-write of the RNGs provided on most machines.
Reproducibility across different machines would also require a reimplementation
of basic transcendental functions normally implemented in hardware by the CPU:
trigonometric, logarithmic, and exponential functions that are commonly used in
simulation codes.  Finally, if a code were to be reproducible for runs with
different numbers of threads/processes, most domain decomposition and load
balancing algorithms would require significant redesign.  Many of these changes
could potentially reduce the speed and efficiency of simulation codes.  As we
have shown here, this effort would simply aid in debugging, without actually
reducing the uncertainty introduced by stochastic variations pumped by stellar
feedback and N-body/hydrodynamic chaos.

\subsection{The physical nature of stochasticity in galaxy formation}
Even if we were to eliminate all uncertainty from our models and purge numerical
artifacts from our codes, the fact would remain that infinitesimally small
changes in initial conditions may still produce macroscopic changes in the final
state of our galaxies. For studies in which one hopes to discover the effects
of different models or parameters for subgrid physics, slightly perturbed
initial conditions have the potential to yield quite different results.

Macroscopic chaotic behaviour makes interpreting the underlying causal
relationship between models for galaxy evolution and the quantities one can
measure difficult, regardless of initial conditions. Comparing the effects of
different physical processes with identical initial conditions, even if
simulated with some hypothetical, perfectly reproducible simulation code will
still have sources for perturbations: those different physical processes. These
will introduce small-scale, irrelevant changes (ones that can be amplified by
the chaotic N-body and hydrodynamic equations), and large-scale, interpretable
changes (the ones we hope to study). Is a decrease in bulge mass an effect of
more efficient feedback \citep{Keller2015}, or a stochastic variation that has
grown into a bar \citep{Sellwood2009}? Is a more metal-rich CGM the effect of
strong outflows \citep{Shen2010}, or a slight change in merger timing? Answering
these sorts of questions with confidence requires some effort to be made in
quantifying the magnitude of stochastic variations on the values being measured.

It is important to keep in mind that the numerical effects explored in this
study are merely the seeds for chaotic behaviour:  RNG seeds and floating-point
roundoff introduce only tiny, infinitesimal variations.  The magnitude of the
variations we see here are much greater than these initial seed perturbations,
because the equations governing the evolution of galaxies have chaotic
solutions.  Both gravitational N-body interactions and hydrodynamic turbulence
have well-known chaotic behaviour, and galaxies are self-gravitating, turbulent
systems.  The coupling of scales introduced by stellar feedback can, as we have
seen, also act to amplify small-scale variations by converting them into
larger-scale injections of energy and momentum.  

While this paper was under review, a related study was published by
\citet{Genel2018}.  This work looks to quantify the same effects we examine here
using a somewhat different strategy.  Rather than looking at individual
galaxies, \citet{Genel2018} uses a series of cosmological volumes to look at
population-scale differences.  This allows them to run a smaller number of
simulations (2-4 sets, depending on their choice of parameters), while still
getting reasonable statistics on the magnitude of stochastic variation.
They use the {\sc AREPO} \citep{Springel2010} hydrodynamics
code, which has been carefully designed to avoid the numerical seed
perturbations we have examined here.  Instead, \citet{Genel2018} explicitly
introduces perturbations in the positions of particles in the initial conditions
on the order of floating point roundoff.  On the scale of their $50^3\Mpc$
simulation volume, this had the effect of adding to each particle a random
position perturbation of $\sim10^{-7}\pc$, roughly 5 solar radii.  Despite
having a completely different source for seed perturbations, \citet{Genel2018}
found similar effects to those we present here.  Critically, they found that
without feedback, increasing resolution decreases the variation in global
properties seen between their pairs of ``shadow'' simulations, which is
consistent with the variation being set by Poisson noise.  However, with the
IllustrisTNG feedback model \citep{Pillepich2018}, stochastic variations became
independent of resolution.  For one of the quantities measured, peak circular
velocity, they found a distribution of variations with a standard deviation
$\sigma\sim0.05$, remarkably similar to the distribution we find for stellar
masses.  Together, our study and \citet{Genel2018} point towards an interesting
new phenomenon: the pumping of stochastic variation through stellar feedback.
Coupling of large \& small scales are a frequent feature of chaotic systems, and
it appears that stellar feedback is another route by which this coupling can be
established.

What these results require is a new way of understanding what numerical
simulations {\it are}.  Rather than mapping a point in the configuration space
of possible initial conditions to a point in an equal-dimensional space of final
states, numerical simulations sample a point within a {\it volume} of this
configuration space, and map it to a point within a different volume.  The size
of this initial volume is constrained by the range of possible initial
conditions that match a constraint we are concerned with: featuring a late major
merger, forming a cluster-mass halo, etc.  The size of the final volume is set
by both the initial conditions and the physical processes involved in our
models.  Small perturbations, introduced either explicitly or implicitly through
numerical effects, are a way to probe the size and shape of this volume.

\section{Conclusions}
We have shown that the small-scale dynamical chaos, seeded by numerical roundoff
and random number generators, can result in significant, long-lasting deviations
in the star formation history and morphology of both isolated and cosmological
galaxies. This fact has a number of implications:
\begin{itemize}
	\item Small differences between simulations generated with different
		physical or numerical models may actually be due to this
		large-scale stochasticity. 
    \item Attributing small ($<20\%$) variations in averaged star formation
        rates {\it absolutely requires} statistical evidence that this is not
        due to stochasticity. This means either running multiple simulations, or
        simulating large volumes which contain many objects. Other quantities
        are also quantitatively \& qualitatively subject to stochastic
        variation.
    \item Stochastic variation is important regardless of whether or not a code
        produces bitwise identical outputs.  Insignificant changes to
        initial conditions or parameters can produce exactly the same effect,
        obscuring the causal relation between changes seen while comparing
        simulations.
    \item Gas exhaustion and self-regulation by feedback can limit stochastic
        deviations, but often require $>\Gyr$ of relatively unperturbed
        evolution to do so. Multiple mergers can push systems to diverge for
        most of the lifetime of the universe.
    \item Stellar feedback efficiently couples different scales within the
        galaxy, and can result in an approximately constant level of variation
        $(\sigma\sim5\%)$ well above what is expected from Poisson noise alone.
	\item  Rapidly varying quantities are more sensitive to stochastic
		variations than smoothed or integrated ones.  Often the
		stochasticity in time is comparable to inter-run variation.
    \item  A single simulation samples a point in a volume of configuration
        space set by the initial conditions and physical model. Numerical
        nondeterminism simply changes where in that volume the final state of
        the simulation ends.
\end{itemize}

We hope that readers do not interpret these results as a Jeremiad on the
prospect of numerical simulation, but instead an exploration of a physical
reality:  galaxies are systems governed by evolution equations with Lyapunov
times shorter than their lifetimes. This does not mean that numerical
simulations are useless, or even that we must always run expensive suites of
simulations in order to generate ensemble averages for all quantities we are
interested in.  Indeed, it is encouraging that our numerical models are now
close enough to nature and each other that distinguishing between them is
becoming non-trivial.

\section*{Acknowledgements}
The analysis was performed using yt (\texttt{http://yt-project.org},
\citealt{yt}) and pynbody (\texttt{http://pynbody.github.io/},
\citealt{pynbody}). We thank Sam Geen and Bernhard R{\"o}ttgers for useful
conversations regarding this paper. We especially would like to thank Oscar
Agertz and Romain Teyssier for providing the {\sc RAMSES} AGORA ICs. The
simulations were performed on the clusters hosted on \textsc{scinet}, part of
ComputeCanada. We greatly appreciate the contributions of these computing
allocations. We also thank NSERC for funding supporting this research. BWK and
JMDK gratefully acknowledge funding from the European Research Council (ERC)
under the European Union's Horizon 2020 research and innovation programme via
the ERC Starting Grant MUSTANG (grant agreement number 714907).
BWK acknowledges funding in the form of a Postdoctoral Research Fellowship from
the Alexander von Humboldt Stiftung. JMDK acknowledges funding from the German
Research Foundation (DFG) in the form of an Emmy Noether Research Group (grant
number KR4801/1-1).
\bibliographystyle{mnras}
\bibliography{references}

\end{document}